\documentclass{IEEEtran}
\PassOptionsToPackage{graphicx}{dvips}
\PassOptionsToPackage{xcolor}{usenames, dvipsnames, table}
\usepackage[scaled=0.88]{helvet}

\usepackage{etex}
\usepackage{makeidx}
\usepackage[lined,linesnumbered,ruled,noend]{algorithm2e}

\usepackage[noend]{algpseudocode}
\usepackage{colortbl}
\definecolor{lgray}{cmyk}{0,00,0,0.3}
\usepackage{amsmath,amsfonts,amssymb}
\usepackage[mathscr]{eucal}
\usepackage{bm}
\usepackage{array}
\usepackage{url}
\usepackage{calc}
\usepackage{float}
\usepackage{latexsym}
\usepackage[hidelinks]{hyperref}
\usepackage{orcidlink}
\usepackage{comment}
\usepackage{multirow}
\usepackage{booktabs}
\usepackage{tabularx}
\newcolumntype{L}[1]{>{\hsize=#1\hsize\raggedright\arraybackslash}X}%
\newcolumntype{R}[1]{>{\hsize=#1\hsize\raggedleft\arraybackslash}X}%
\newcolumntype{C}[1]{>{\hsize=#1\hsize\centering\arraybackslash}X}%
\usepackage{pdfpages}
\usepackage{tabu}
\DeclareGraphicsExtensions{.eps,.jpg,.png,.pdf}
\usepackage{multicol}
\usepackage{tikz} 
\usetikzlibrary{positioning,chains,fit,shapes,calc}

\usepackage{todonotes}
\usepackage{xspace}
\usepackage{caption}
\usepackage[export]{adjustbox}
\newcommand{\ie}{\textit{i.e.}\xspace}

\newcommand\blfootnote[1]{%
	\begingroup
	\renewcommand\thefootnote{}\footnote{#1}%
	\addtocounter{footnote}{-1}%
	\endgroup
}

\newtheorem{example}{Example}
\newtheorem{remark}{Example}

\AtBeginEnvironment{quote}{\sf}

\let\oldtabular=\tabular
\def\tabular{\small\oldtabular}

\begin{document} 
	
\abovedisplayskip=0.1mm

\title{A New Tool to Find Lightweight (AND, XOR) Implementations of Quadratic Vectorial Boolean Functions up to Dimension 9}
\author{Marie Bolzer\orcidlink{0009-0009-2601-1684}, S\'{e}bastien Duval\orcidlink{0000-0003-2558-1756}, and Marine Minier\orcidlink{0000-0003-3252-2578}
\thanks{This work was partially supported by the French National Agency of Research under grant number ANR-22-CE39-0015.}
\thanks{Marie Bolzer, S\'{e}bastien Duval and Marine Minier are with Universit\'e de Lorraine, CNRS, Loria, Inria, France (e-mails: \href{mailto:marie.bolzer@loria.fr}{marie.bolzer@loria.fr}, \href{mailto:sebastien.duval@loria.fr}{sebastien.duval@loria.fr}, \href{mailto:marine.minier@loria.fr}{marine.minier@loria.fr}).}
}
\maketitle               

\begin{abstract}
The problem of finding a minimal circuit to implement a given function is one of the oldest in electronics. It is known to be NP-hard. Still, many tools exist to find sub-optimal circuits to implement a function. In electronics, such tools are known as synthesisers. However, these synthesisers aim to implement very large functions (a whole electronic chip). In cryptography, the focus is on small functions, hence the necessity for new dedicated tools for small functions.

Several tools exist to implement small functions. They differ by their algorithmic approach (some are based on Depth-First-Search as introduced by Ullrich in 2011, some are based on SAT-solvers like the tool desgined by Stoffelen in 2016, some non-generic tools use subfield decomposition) and by their optimisation criteria (some optimise for circuit size, others for circuit depth, and some for side-channel-protected implementations). However, these tools are limited to functions operating on less than 5 bits, sometimes 6 bits for quadratic functions, or to very simple functions. The limitation lies in a high computing time.

We propose a new tool\footnote{The tool is provided alongside the IEEE article with CodeOcean and at \url{https://github.com/seduval/implem-quad-sbox}.}
to implement quadratic functions up to 9 bits within AND-depth 1, minimising the number of AND gates. This tool is more time-efficient than previous ones, allowing to explore larger implementations than others on 6 bits or less and allows to reach larger sizes, up to 9 bits.
\end{abstract} 

\blfootnote{This work has been submitted to the IEEE for possible publication. Copyright may be transferred without notice, after which this version may no longer be accessible.}

\IEEEpubid{}

\section{Introduction}\label{sec:intro}
\noindent

\subsection{Motivation}
Finding lightweight implementations is important in any field, but cryptography has a slightly different setting compared to other fields. In particular, cryptography needs to be efficient both in software and hardware, both on servers and on small on-chip systems. Cryptography is a costly overhead in everyday communications, hence lightweight cryptography has received a lot of attention for decades.

A specificity in the field of cryptography is that the functions that need to be implemented are well-known, well-studied, and designed to reduce costs. In particular, we know that the bulk of the circuit cost lies in small functions which are repeated many times. These functions, called S-boxes, typically operate on 3 to 8 bits. Another specificity is that by default, the implementation of cryptographic functions must be protected against side-channel attacks, which makes binary XOR gates (and any linear gate) much cheaper than AND gates, unlike a more classical setting. The S-boxes being the only non-linear parts of block ciphers and being repeated many times, they are the main focus when optimising cryptographic functions.

Specifically, protecting linear gates like XOR is cheap, while protecting non-linear gates like AND is expensive~\cite{DBLP:conf/crypto/IshaiSW03}. For this reason, the main goal in S-box optimisation is to reduce the number of AND gates. A secondary objective is to reduce the AND depth, which determines the latency of hardware protected implementations~\cite{bilginthesis}. Finally, reducing the number of XOR gates is less crucial but still has some impact on the cost.

Tools for optimising the implementation of S-boxes are of two types: (i) tools which generate electronic circuits until the corresponding S-box has the desired mathematical properties and (ii) tools which search for the best circuit to implement a given S-box.
These two approaches are linked: (ii) is a sub-case of (i) in which the desired property is equality to the given S-box. Also, applying the tool for (ii) on all S-boxes which verify the desired properties gives a tool for (i) (it is easier using S-boxes classification~\cite{DBLP:conf/eurocrypt/BiryukovCBP03}).

Although the current trend is mostly to use small S-boxes (on 3 or 4 bits), we argue that larger S-boxes are more worthwhile. Indeed, (i) larger S-boxes allow for stronger non-linear layers (with multiple S-boxes in parallel), (ii) the security reduction of block ciphers to S-boxes using usual techniques (such as the wide-trail strategy~\cite{DBLP:conf/ima/DaemenR01}) depends on less approximation for larger S-boxes and (iii) larger S-boxes may allow for better trade-offs between cost and security~\cite{DBLP:journals/tosc/BilginMDLS20}. The main drawback to using larger S-boxes is actually that they have costly implementations because current tools for complex S-boxes give sub-optimal results.
Note also that against the trend, the most recent symmetric NIST standards, Keccak~\cite{DBLP:conf/dagstuhl/BertoniDPA09a} and ASCON~\cite{DBLP:journals/joc/DobraunigEMS21}, use 5-bit S-boxes.

\paragraph{Previous tools}
A few previous tools exist, mostly limited to the implementation of 4-bit permutations~\cite{ullrich2011finding,Gla,DBLP:journals/joc/BoyarMP13,DBLP:journals/tosc/JeanPST17,DBLP:journals/tosc/BaoGLS19}. One notable exception is the tool by Stoffelen~\cite{DBLP:conf/fse/Stoffelen16} which works on few 5-bit functions, a tool which was enhanced in~\cite{DBLP:journals/tosc/BilginMDLS20} to work on all 5-bit functions and some 6-bit quadratic functions. This tool is based on a black-box SAT solver and thus allows for little improvement. The SAT model was recently refined in~\cite{DBLP:journals/iacr/ZhangH23} and in~\cite{DBLP:journals/tcasI/FengWZPZ24} with similar results but better speed, and the ability to consider very simple S-boxes (like the Keccak $\chi$ mapping) on up to 8 bits.

On top of previous tools for S-box implementation, there is a series of tools to solve the problem of classification according to affine equivalence, which has similarities with synthesisers. Actually, our tool resembles the affine equivalence algorithm by Biryukov~\emph{et al.}~\cite{DBLP:conf/eurocrypt/BiryukovCBP03}, which boils down to solving a large linear system. One notable such tool is the one by De Meyer and Bilgin~\cite{DBLP:journals/iacr/MeyerB18}, focused on quadratic functions and combined with quadratic decomposition to reach larger degrees.

\subsection{High-level considerations on the problem}
S-box implementation has been studied for at least 30 years, but large S-boxes remain hard to implement. We propose to take a step back and reconsider the problem itself.

\paragraph{Problem definition}
One is given an S-box, represented by its look-up-table (LUT) or by its algebraic normal form (ANF). The goal is to output a circuit which implements the S-box optimally given a cost metric (Gate-Equivalent, number of AND, circuit depth, \ldots).

\paragraph{Problem hardness}
How hard is this problem really? We know it is NP-hard~\cite{DBLP:journals/joc/BoyarMP13}, but this only concerns infinitely large S-boxes. How hard is it for 4-bit S-boxes? For 10-bit S-boxes? To estimate this hardness, we attempted a little experiment: take the ANF of an S-box and find a lightweight implementation with pen, paper and coffee.

We experimented with all S-boxes from~\cite{DBLP:journals/tosc/BilginMDLS20} (\emph{i.e.}~4- and 5-bit S-boxes, and quadratic 6-bit S-boxes), optimising for AND gates and AND depth. To our surprise, we obtained as good implementations by hand as with tools. What is more, it took us around 4 hours by hand per implementation, similar to the tool's speed (although better models~\cite{DBLP:journals/iacr/ZhangH23, DBLP:journals/tcasI/FengWZPZ24} run within seconds). It becomes impractical for 7-bit S-boxes.

It appears that for practical sizes, the problem is actually not \emph{that} hard. Thus rose a question: Computers are powerful tools, why are they not more efficient for this problem?

\paragraph{Algorithmic considerations}
We think it is because of algorithmic choices. The first tools used ad-hoc algorithms~\cite{ullrich2011finding}: build random circuits until a circuit matches the S-box. It is highly random within a gigantic search space. More recent tools use SAT-solvers, which depend on a good generic mathematical model, but such a generic model is hard to define (though some steps in this direction are reached in~\cite{DBLP:journals/tcasI/FengWZPZ24}).

\paragraph{Pen-and-paper algorithm}
However the pen-and-paper algorithm that seems natural to us works very differently: starting from the ANF, we look for patterns which appear in several output bits and iteratively factorise the ANF. In principle, this explains why computers are not efficient for this algorithm: the human brain is efficient at pattern-matching, but computers are not because they are sequential.

\paragraph{Research directions}
Based on these observations, we think several research directions are valuable: (i) finding a better mathematical representation of the problem to model it better~\cite{DBLP:journals/iacr/ZhangH23, DBLP:journals/tcasI/FengWZPZ24}, (ii) finding new algorithms more adapted to computers than random search (which can be combined with (i)), and (iii) attempting to implement pattern-matching using computing tricks.

In this work, we tackle the third approach. Several ideas can come to mind to implement pattern-matching: (a) machine-learning techniques have been tuned for this purpose but seem costly, (b) making use of massive parallelism like in GPU could ease pattern-matching and (c) trying to precompute the pattern-matching effort.

We chose to study problem (iii) with the approach (c) and got some new promising results. Note however that many new valuable directions arise for future works. Also, we started with quadratic S-boxes because for larger degrees, precomputing all the patterns is a hard problem in itself (we plan to tackle this problem in a future work).

\subsection{Our setting}
Alike~\cite{DBLP:journals/tosc/BilginMDLS20}, we focus on reducing the number of AND gates and the AND depth to reduce the cost of secure implementations.
Indeed, since the 1990s~\cite{DBLP:conf/crypto/ChariJRR99,DBLP:conf/ches/GoubinP99,DBLP:conf/tcc/MicaliR04}, we know that cryptographic implementations must be protected by default\footnote{For example, the NIST call for Lightweight Cryptography included ``The implementations of the AEAD algorithms and the optional hash function algorithms should lend themselves to countermeasures against various side-channel attacks''.}. The usual protection is masking~\cite{DBLP:journals/cacm/Shamir79,DBLP:conf/mark2/Blakley79}, which separates each secret in $d$ shares, which implies very costly circuits for AND gates, and a slowdown linear in the AND depth~\cite{DBLP:conf/crypto/IshaiSW03,DBLP:journals/jce/BozilovKN22}.

Originally reaching security order $t$ required $d=2t+1$ shares~\cite{DBLP:conf/crypto/IshaiSW03}, but since~\cite{DBLP:conf/ccs/BartheBDFGSZ16,DBLP:journals/tches/GoudarziPRV21} we know it is doable with $d=t+1$ shares, and that high-order masking is necessary (the authors consider $d$ up to 100). From~\cite{DBLP:conf/crypto/IshaiSW03}, we also know that, in theory, protecting an AND gate costs a factor $\mathcal{O}(d^2)$ while for a XOR gate it is $\mathcal{O}(d)$.
It remains to know for which value $t$ an implementation is secure. Challenges have been proposed to solve this question: From~\cite{DBLP:journals/iacr/BronchainCS21}, we know that $d=2$ is breakable in 5 minutes with a single trace. From the Spook challenge\footnote{\url{spook.dev}}~\cite{DBLP:journals/tosc/BelliziaBBCDGLL20} we know that $d=8$ is breakable with $70,000 \sim 2^{16}$ traces, still far from secure. From~\cite{DBLP:conf/ches/JournaultS17} we know that $d=32$ is secure. Roughly, it seems that it takes $2^{2d}$ traces to attack $d$ shares, indicating that $d>20$ is required, and $d=32$ seems reasonable to reach practical security with enough margin.
Putting things together, an AND gate costs a factor roughly $32^2=1024$ to be masked in theory. Practical tests exist. In~\cite{DBLP:conf/eurocrypt/GoudarziR17}, the authors study timings: they show that a very optimised AES S-box layer with $d=10$ costs $300,000$ cycles, while a linear layer costs 315 cycles, on an ARM 32-bit processor (typical of small industrial objects), and the whole AES~\cite{DBLP:books/sp/DaemenR02} costs $1906.5 d^2 + 10972.5 d + 7712$. For PRESENT~\cite{DBLP:conf/ches/BogdanovKLPPRSV07}, the S-box layer costs $30,000$ cycles, the linear one 160 cycles. Note that this is with $d=10$, so not even secure. On larger processors,~\cite{DBLP:conf/ccs/BartheBDFGSZ16} shows that AES with $d=20$ takes 1min on an Intel Xeon, 769 times slower than an unprotected implementation, and Keccak~\cite{DBLP:conf/dagstuhl/BertoniDPA09a} takes 3min.

From this, it appears that lightweight secure implementations need to minimise the number of AND gates and AND depth, while the number of XOR gates and XOR depth have a negligble impact.

In order to minimise the AND depth, we limit our implementations to those with AND depth equal to 1 which is optimal for quadratic functions. It remains to reduce the number of AND gates, which is the focus of our tool.

We focus on circuits using binary (AND, XOR) gates\footnote{Note that this setting implicitly includes the NOT gate as NOT $x = x$ XOR 1.}. We limit to binary gates as we aim for bit-sliced implementations which allow for binary-gate parallelism. We limit to AND and XOR as it allows to have a non-linear gate (AND), a linear gate (XOR), and gives the property that any binary non-linear gate can be implemented with a single AND gate (\textit{i.e.} any other gate can be obtained without wasting any non-linear gate).

Our tool is limited to quadratic S-boxes. This is due to a high memory complexity when considering large (more than 6-bit) S-boxes with degree more than 2, but more than quadratic is a valuable goal which we aim to tackle in the future. Note however that efficient implementations of quadratic S-boxes is already a valuable achievement, as many quadratic S-boxes are used in practice. This is notably the case of the two most recent NIST standards Keccak~\cite{DBLP:conf/dagstuhl/BertoniDPA09a} and ASCON~\cite{DBLP:journals/joc/DobraunigEMS21}, which are quadratic on 5 bits. 

\paragraph{Organisation}
This paper is organised as follows: Section \ref{sec:fomalising} formalises the problem with small examples ; Section~\ref{sec:tool} presents our algorithm and its embedding in our tool ; Section~\ref{sec:results} gives the results we obtained for the quadratic case and compares our timings with previous works. Finally, Section~\ref{sec:conclusions} gives a conclusion.

\section{Formalising the Problem}\label{sec:fomalising}
\noindent

Our main idea is to mimic what we do by hand and decompose the problem.
Computers are inherently bad at finding patterns, hence we precompute and store all the patterns. Then finding an implementation of S-box $S$ is reduced to finding, for each output bit $y_i$, a linear combination of some stored patterns which equals to $y_i$.

\begin{example}
	Let $y$ be one of the output bits we want to implement, $y = x_0x_1 \oplus x_1x_2 \oplus x_2x_3$. Suppose we have stored patterns $q_0 = x_0x_1 \oplus x_1x_2 = x_1(x_0 \oplus x_2)$, $q_1 = x_1x_2 \oplus x_2x_3 = x_2(x_1 \oplus x_3)$, $q_2 = x_0x_1$, $q_3 = x_1x_2$ and $q_4 = x_2x_3$, then our problem is reduced to solving a linear system in the variables $q_i$. Let $Q=(q_0,q_1,q_2,q_3,q_4)^T$. We are looking for a vector $I$ such as $I \times Q
	= y.$
An example solution is $I= (1 0 0 0 1)$ which gives $y=q_0 \oplus q_4$.
\end{example}

The first problem we hit is the number of patterns to store. To reduce it, we separate our problem in two parts: linear and non-linear. Indeed, linear operations only affect the number of XOR gates, which is a secondary objective. We optimise the non-linear part first, minimising the number of AND gates. This step outputs a list of linear functions required to implement the non-linear part. In a second step, we take these linear functions and find an implementation of all of them jointly, reducing the number of XOR gates.

The input of the non-linear part will consist of the variables truncated from their linear parts: we truncate the linear parts from the patterns $q_i$ and from the output bits $y_i$.

\begin{example}[Illustration with $n=3$]
	
	Consider the S-box with ANF:
	\begin{flalign*}
	\hspace{\parindent} y_0 &= x_0 \oplus x_1 \oplus x_2 \oplus x_1x_2&\\
	\hspace{\parindent} y_1 &= x_1 \oplus x_0x_1 \oplus x_0x_2&\\
	\hspace{\parindent} y_2 &= x_0x_1 \oplus x_2&
	\end{flalign*}
	Listing all non-linear operations truncated from their linear parts gives matrix $Q$:

	\nopagebreak
	\begin{minipage}{.3\textwidth}\small
	\begin{align*}
	&x_0x_1, \\[0pt]
	&x_0x_2, \\[0pt]
	&x_1x_2, \\[0pt]
	&x_0x_1 \oplus x_0x_2, \text{given for example by:~} x_0(x_1 \oplus x_2) \\[0pt]
	&x_0x_1 \oplus x_1x_2, \text{given for example by:~} x_1(x_0 \oplus x_2) \\[0pt]
	&x_0x_2 \oplus x_1x_2, \text{given for example by:~} x_2(x_0 \oplus x_1) \\[0pt]
	&x_0x_1 \oplus x_0x_2 \oplus x_1x_2, \text{given for example by:~} (x_0 \oplus x_2)(x_0 \oplus x_1)
	\end{align*}
	\end{minipage}\hspace{8pt}
	\begin{minipage}{.25\textwidth}
	$$Q = \begin{pmatrix}
		x_0x_1 \\[3pt]
		x_0x_2 \\[3pt]
		x_1x_2 \\[3pt]
		x_0x_1 \oplus x_0x_2 \\[3pt]
		x_0x_1 \oplus x_1x_2 \\[3pt]
		x_0x_2 \oplus x_1x_2 \\[3pt]
		x_0x_1 \oplus x_0x_2 \oplus x_1x_2
	\end{pmatrix}.$$
	\end{minipage}

	\medskip

	Consider matrix $Y$ representing the ANF, truncated from its linear parts:
	 We have
	 \[Y = \begin{pmatrix}
		x_1x_2\\
		x_0x_1 \oplus x_0x_2 \\
		x_0x_1 \\
	\end{pmatrix}.\]
	Every implementation of the non-linear part is then a solution of the equation
	\(IQ = Y \), where the size of $I$ is (number of output bits) $\times$ (size of $Q$).

	\[I_1 = \begin{pmatrix}
		1 & 0 & 0 & 0 & 1 & 0 & 0 \\
		1 & 1 & 0 & 0 & 0 & 0 & 0 \\
		0 & 1 & 1 & 0 & 0 & 0 & 1 \\
	\end{pmatrix}\]
	is a solution giving implementation:
	\begin{align*}
	q_1 &= x_0x_1  \\
	q_2 &= x_1(x_0 \oplus x_2) \\
	y_0 &= q_1 \oplus q_2 \\
	q_3 &= x_0x_2 \\
	y_1 &= q_1 \oplus q_3 \\
	q_4 &= x_1x_2 \\
	q_5 &= (x_0 \oplus x_2)(x_0 \oplus x_1) \oplus x_0 \\
	y_2 &= q_3 \oplus q_4 \oplus q_5 
	\end{align*}
	\
	\[I_2 = \begin{pmatrix}
		0 & 0 & 1 & 0 & 0 & 0 & 0 \\
		0 & 0 & 0 & 1 & 0 & 0 & 0 \\
		1 & 0 & 0 & 0 & 0 & 0 & 0 \\
	\end{pmatrix}\]
	is another solution giving implementation:
	\smallskip
	\begin{align*}
		q_1 &= x_1x_2  \\
		y_0 &= q_1 \\
		q_2 &= x_0(x_1 \oplus x_2) \\
		y_1 &= q_2 \\
		q_3 &= x_0x_1 \\
		y_2 &= q_3 
	\end{align*}

	A crucial observation is that the number of AND gates matches the number of non-zero columns in $I$. Thus finding the `optimal' implementation means finding the solution with the most zero columns.
	Moving to the linear step, to get the final implementation, we need to add the missing linear expressions.
	For solution $I_2$, an example final implementation is:
	
	\begin{align*}
	 l_0 &= x_1 \oplus x_2 \\
	 l_1 &= x_0 \oplus l_0 \\
	 q_1 &= x_1x_2 \\ 
	 y_0 &= q_1  \oplus l_1 \\ 
	 q_2 &= x_0l_0  \oplus x_1  \\
	 y_1 &= q_2  \\ 
	 q_3 &= x_0x_1  \\
	 y_2 &= q_3 \oplus x_2 
	\end{align*}
	\label{example of formalisation}
	
\end{example}

\paragraph{The number of patterns}
What we called ``patterns'' in a hand-based approach actually consists of all Boolean functions obtainable in a single AND gate.
Despite truncating the linear parts, the number of such 1-AND-gate functions remains large. We computed the exact number for S-boxes from 3 to 11 bits in Column 2 of Table~\ref{table:set_op_size}.
With linear combinations of 1-AND-gate operations, we can implement any function of degree lesser than or equal to 2 (quadratic).
Observe that this number increases very fast. The total execution time depends greatly on this number.

\begin{table}[H]\centering
	\begin{tabular}{c|c}
		\hline
		Number of & Number of quadratic polynomials \\
		bits& (truncated) obtained with 1 AND  \\
		\hline
		3 & 7 \\
		\hline
		4 & 35 \\
		\hline
		5 & 155 \\
		\hline
		6 & 651 \\
		\hline
		7 & 2667 \\
		\hline
		8 & 10 795 \\
		\hline
		9 & 43 435 \\
		\hline
		10 & 172 846 \\
		\hline
		11 & 692 775 \\
		\hline
	\end{tabular} 
	\caption{Number of patterns per size.}
	\label{table:set_op_size}
\end{table}

\section{Description of the Tool}\label{sec:tool}
\noindent

In this section, we describe our tool first from an algorithmic point of view in Section~\ref{sec:algo} and then from an implementation point of view in Section~\ref{sec:Cimplem}.

\subsection{Algorithms used} \label{sec:algo}

\subsubsection{High-level description of the algorithm}

Our tool works in two parts. First we truncate the linear parts from all expressions and optimise the \textit{non-linear} parts, yielding a set of quadratic operations implementing the non-linear parts of the output bits. Then from this set, we generate another set containing the linear operations needed for the \textit{linear} parts. We put it all together to get a circuit for the S-box. The tool output is a C program implementing the S-box.

The non-linear part uses two sets. \textsf{Set\_op}: set of all operations obtainable in exactly 1 AND (matrix $Q$ in Example~\ref{example of formalisation}). \textsf{Op\_Selec}: output of the non-linear part. It starts empty and the algorithm fills it with the quadratic operations needed for the implementation.

We designed an ad-hoc tool to solve the matrix system associated with the non-linear part. The idea is to implement output bits one by one but not independently: when processing the $i$-th bit, we express its Boolean function as a XOR of elements of \textsf{Set\_op}, giving priority to operations already used to implement the previous bits. At the end of step $i$, \textsf{Op\_selec} contains the operations needed to implement the first $i$ output bits.

The core of the algorithm is the way \textsf{Op\_selec} is updated. We look for a XOR-sum of elements of \textsf{Op\_selec} and \textsf{Set\_op} that equals the $i$-th output bit.
Each element of \textsf{Set\_op} in the sum is then added to the updated \textsf{Op\_selec} and adds 1 AND to the circuit. Thus we add as few elements of \textsf{Set\_op} as possible in the sum.

In essence, we make successive local optima, output bit by output bit, to reach a global sub-optimum, which is a greedy algorithm. When it is computationally feasible, we exhaust all successions of local optima, then sub-optima, which allows to reach the global optimum.

It is a branch-and-bound algorithm: we span a tree where level $i$ contains all possible implementations of the $i$-th output bit\footnote{Hence the size of the tree is exponential in the number of output bits.}. We bound on the number of AND gates to cut branches as soon as $\#$\textsf{Op\_selec} $>$ \textsf{nb\_and\_max}, where \textsf{nb\_and\_max} is a bound set by the user. Actually, we cut earlier by computing a lower bound on the number of AND needed to implement the remaining output bits (which gives an $A^*$ heuristic~\cite{DBLP:journals/tssc/HartNR68,moustakas1990heuristic}).

The algorithm for the linear part is similar, except that it is less exhaustive since we are looking for less optimal solutions.

Section~\ref{subsec:nonlin} details the non-linear algorithm (its pseudo-code is given in Algorithm~\ref{Algo-non-lin}). Section~\ref{subsec:lin} details the linear algorithm (its pseudo-code is given in Algorithm~\ref{Algo-lin}).

\subsubsection{Algorithm for the non-linear part} \label{subsec:nonlin}

\paragraph{Description of the algorithm}

 We start with:
\begin{itemize}
	\item \textsf{Set\_op}: set of all operations we can obtain by using exactly 1 AND,
	\item \textsf{Op\_selec}: set of ``selected'' operations: it will contain the operations needed for the implementation. It is updated as the algorithm progresses,
	\item \textsf{nb\_and\_max}: bound on the number of AND gates (to bound the search space).
\end{itemize}
At each stage $1\leq i\leq n$, we add to \textsf{Op\_selec} the non-linear operations needed to implement output bit $y_i$. To minimise the added operations, we test in order if we need to add:
\begin{itemize}
	\item 0 AND gate:
	Corresponding to the case where $y_i$ can be implemented using only the operations already available to us (\ie the \textit{selected} operations in \textsf{Op\_selec}),
	\item 1 AND gate:
	Corresponding to the case where we need to add only 1 operation (from \textsf{Set\_op}) to the set of selected operations to implement $y_i$,
	\item 2 AND gates, etc.
\end{itemize}

In detail, consider \textsf{Op\_selec}$_{i-1}$ the set of operations to implement $y_0,\,\ldots,\,y_{i-1}$, of cardinality $C$. We allow to add at most $m = \textsf{nb\_and\_max} - C$ operations to \textsf{Op\_selec}. We then loop over all $0\leq j\leq m$ and $0\leq k\leq C$ and test if a XOR-sum of $j$ elements of \textsf{Op\_selec}$_{i-1}$  and $k$ elements of \textsf{Set\_op} equals $y_i$. We obtain the updated \textsf{Op\_selec}$_i$ by adding the $k$ elements of \textsf{Set\_op} to \textsf{Op\_selec}$_{i-1}$, thus the cost of adding $y_i$ is $k$ AND gates. The function looking for XOR-sums equal to $y_i$ is hereafter denoted $\textbf{\textsf{test\_xor}}(\textsf{Op\_selec},y_i,j,k)$.

\paragraph{Multiplicity of added operations}

First, note that the set of operations returned by a loop is not unique. 

\begin{example}
	Consider the ANF:
	\begin{align*}
		y_0 &= x_0x_2 \\
		y_1 &= x_0x_1 \oplus x_2x_3\\
		y_2 &= x_0x_3 \oplus x_1x_2 \oplus x_2x_3
	\end{align*}
	
	Assume we have already dealt with $y_0$ and that the algorithm has selected the following operation:  
	\[ \textsf{Op\_selec} = \{ q_1 = x_0x_2 \} \]
	We cannot obtain $y_1$ only using \textsf{Op\_selec} and we cannot obtain it by adding only one operation. But we can obtain it by adding two operations and we have several options:
	\begin{itemize}
		\item We can add $q_2 = x_0x_1$ and $q_3 =  x_2x_3$ to \textsf{Op\_selec}. \\
		Then, we have $y_1 = q_2 \oplus q_3,\ \textsf{Op\_selec} = \{ q_1, q_2, q_3 \}.$
		\item We can add $q_4 = x_0x_1 \oplus x_1x_2$ and $q_5 = x_1x_2 \oplus x_2x_3$ to \textsf{Op\_selec}. \\
		Then, we have $y_2 = q_4 \oplus q_5,\ \textsf{Op\_selec} = \{ q_1, q_4, q_5 \}. $
	\end{itemize}
	
	At this stage, we cannot know which one of the options will be the best, so we need to store all the options and test each one if we want to find an optimal solution.
	In this case, the second option will allow us to obtain $y_2$ by adding only one operation ($x_0x_3$) which is not the case with the first solution, so the second option will end up best.
\end{example}

\paragraph{Impact of the order of treatment of the output bits}

The order in which the output bits are treated has an impact on whether or not a solution is obtained.

\begin{example}
	Suppose that we run our algorithm on an example S-box, with the order ($y_3$ - $y_0$ - $y_2$ - $y_1$) for the output bits, and obtain the resulting implementation with 7 non-linear operations $q_i$:
	\begin{align*}
		 y_3 &= q_0 \oplus q_4 \oplus q_5 \\
		 y_0 &= q_5 \oplus q_6 \\
		 y_2 &= q_2 \oplus q_3 \\
		 y_1 &= q_0 \oplus q_1
	\end{align*}
	
	Now, assume that bit $y_0$ can also be obtained by doing $q_1 \oplus q_3 \oplus q_4$ and that instead we use the order $(y_1$ - $y_2$ - $y_0$ - $y_3)$. In the step to obtain $y_0$, since the algorithm will be looking for the least expensive implementation, it will return $y_0 = q_1 \oplus q_3 \oplus q_4$, which requires the addition of just 1 non-linear operation, since $q_1$ and $q_3$ are for free as they have already been used previously to implement $y_1$ and $y_2$. The solution $y_0 = q_5 \oplus q_6$ requires 2 additional non-linear gates, as neither $q_5$ or $q_6$ are for free so it will not be chosen. The final implementation will then include $6$ non-linear operations:
	\begin{align*}
		y_1 &= q_0 \oplus q_1 \\
		y_2 &= q_2 \oplus q_3 \\
		y_0 &= q_1 \oplus q_3 \oplus q_4 \\
		y_3 &= q_0 \oplus q_4 \oplus q_5
	\end{align*}
\end{example}
So, if we want to be sure of getting the best implementation, we need to try all the possible permutations of the output bits.
This is expensive: for $n=10$, there are $10!=3628800$ permutations of the output bits, so this implies a factor 3628800 on the number of operations to run the algorithm. Implementations can be obtained much faster by not considering all permutations, but we do not have a good way of knowing in advance which permutations will be good or bad for implementation-cost, so this would yield sub-optimal results.

In order to reduce this cost, we have chosen to start the algorithm with an initialised \textsf{Op\_selec}. To do this, we look for the ``cheapest'' output bit, \textit{i.e.} the output bit requiring the fewest non-linear operations. The different ways of implementing this bit will correspond to the initialisation of \textsf{Op\_selec}. This allows us to test $(n-1)!$ permutations instead of $n!$ and also facilitates code parallelisation.

\paragraph{Algorithm}
The optimisation of the non-linear part is summarised in Algorithm~\ref{Algo-non-lin}. (For readability reasons, the return of $\textsf{test\_xor}$ is considered unique).

\begin{algorithm}[]
	\caption{Optimisation of the non-linear part.}
	\label{Algo-non-lin}
	\SetKwInOut{Input}{Input}\SetKwInOut{Output}{Output}
	\Input{
		\begin{itemize}
			\item $\{y_0,....,y_{n-1}\}$: output bits of the S-box, truncated from linear parts,
			\item \textsf{Set\_op}: fixed set of all operations obtainable in 1 AND,
			\item \textsf{Op\_selec}: set of operations, initially empty (its size is denoted $C$),
			\item \textsf{nb\_and\_max}: integer, bound on number of AND gates given by user.
			\end{itemize}
	}
	
	\Output{
		\begin{itemize}
			\item \textsf{Op\_selec}: set of operations required to implement the S-box.
		\end{itemize}
	}

	$y_k$ $\leftarrow$ cheapest output bit \\
	\For {All initialisation sets $S$ to obtain $y_k$}
	{\textsf{Op\_selec} $\leftarrow$ $S$\\
		\For {Each permutation of $\{y_0,....,y_{n-1}\} \setminus \{y_k\}$ }
		{
			$\textsf{nb\_and} \leftarrow \textsf{nb\_and\_max - S.size()}$ \\
			\For {$i$ from $0$ to $n-1$} 
			{\tcp{Browse all output bits}
				\For {$j$ from $0$ to $\textsf{nb\_and}$}
				{\tcp{Add as few non-linear operations as possible}
					\For {$k$ from $0$ to $C$}
					{\tcp{Minimise the total number of operations used}
						$S = \textsf{test\_xor} (\textsf{Op\_selec},y_i,k,j)$ \tcp{Test if there are linear combinations of $j$ elements of \textsf{Set\_op} and $k$ elements of \textsf{Op\_selec} equal to $y_i$}
						\If {$ S = \emptyset$}
						{
							\textbf{continue}
						}
						\Else { 
							$\textsf{nb\_and} \leftarrow \textsf{nb\_and} - j$ \tcp{Update the AND-gate count}
							$ \textsf{Op\_selec} \leftarrow \textsf{Op\_selec}\ \cup\ S$ \tcp{Update the implementation}
							\textbf{goto (6)} \tcp{Go back to line 6, move to the next $y_k$}
						}
					}
					\textbf{continue} \tcp{No solution with $j$ from \textsf{Set\_op} $\rightarrow$ try $j+1$}
				}
				\textbf{break} \tcp {No solution $\rightarrow$ try the next permutation}
			}
			\Return \textsf{Op\_selec}
		}
	}
\end{algorithm}

\subsubsection{Algorithm for the linear part}\label{subsec:lin}

The aim of this second step is to implement all the linear expressions we will need. This step builds on the result of the previous step as  the set of required linear expressions consists of: 
\begin{enumerate}
	\item truncated parts of the ANFs, 
	\item linear operations to obtain the non-linear operations calculated in the previous step,
	\item truncated parts of the non-linear operations.
\end{enumerate}

\begin{example}
	Consider $q = x_0x_1 \oplus x_1x_2 \oplus x_0x_2$, we cannot get it on its own: it is not obtainable in only 1 AND gate. It can be obtained, for example, by doing $(x_0 \oplus x_1)(x_1 \oplus x_2)$, which gives $x_0x_1 \oplus x_0x_2 \oplus x_1 \oplus x_1x_2$. We therefore need to add $x_1$ as a type-3 linear expression, and we require $x_0 \oplus x_1$ and $x_1 \oplus x_2$ as type-2 linear expressions.
\end{example}

Once this set is constructed, the principle is basically the same as in the previous step. We process the expressions one after the other and increment a set of selected operations.

\paragraph{Description of the algorithm}
We start with:
\begin{itemize}
	\item $L$: set of linear expressions $l_i$ required,
	\item \textsf{Op\_selec}: set of ``selected'' operations (like in the previous algorithm), initialised with all the linear monomials $x_i$. It will contain the linear operations required for the implementation and is incremented as the algorithm progresses.
	\item \textsf{nb\_xor\_max}: bound on the number of XOR gates required.
\end{itemize}

In detail, consider \textsf{Op\_selec}$_{i-1}$ the set of operations to implement $l_0,\,\ldots,\,l_{i-1}$, of cardinality $C$. We allow to add at most $m = \textsf{nb\_xor\_max} - C$ operations to \textsf{Op\_selec}. We then loop over all $0\leq j\leq m$ and test whether there is a XOR-sum of $j$ elements of \textsf{Op\_selec} equal to $l_i$. We add to \textsf{Op\_selec} the element $l_i$, as well as an arbitrary set of intermediate variables to obtain $l_i$ as the sum of the $j$ elements (see the following Examples). Overall, \textsf{Op\_selec}$_i$ costs $j$ more XOR than \textsf{Op\_selec}$_{i-1}$.

\paragraph{Multiplicity of decompositions of a non-linear expression}

Note that there are several sets of linear operations whose AND corresponds to a given non-linear operation.
\begin{example}
	Consider the non-linear operation $q_1 = x_1x_2 \oplus x_1x_3 \oplus x_2x_3$. \\
	The set of linear operations to implement $q_1$ are as follows:
	\begin{align*}
		 x_1 \oplus x_2 \text{~,~} x_1 \oplus x_3 \text{~(as type 2)~} \text{~and~} x_1 \text{~(as type 3)}, \\
		 x_1 \oplus x_2 \text{~,~} x_2 \oplus x_3 \text{~(as type 2)~} \text{~and~} x_2 \text{~(as type 3)}, \\
		 x_1 \oplus x_3 \text{~,~} x_2 \oplus x_3 \text{~(as type 2)~} \text{~and~} x_3 \text{~(as type 3)}.
	\end{align*} 
\end{example}
Finding the optimal implementation requires to test all options to know which one is the least expensive in the end. But the number of cases increases exponentially, and linear optimality is not one of our goals, hence we decide to pick only one option arbitrarily\footnote{\label{footnoteNOT}This arbitrary choice never picks a linear expression involving a $\oplus 1$ (NOT gate).}.

\paragraph{Multiplicity in linear decompositions}
Similarly, a linear expression can be decomposed in many ways, meaning that there are several ways to increment \textsf{Op\_selec}.

\begin{example}
	Consider the linear expression $l = x_0 \oplus x_1 \oplus x_2 \oplus x_3$ and assume we have $l_0 = x_0 \oplus x_2$ in addition to the linear monomials ($x_0, x_1, x_2, x_3)$ in the set of selected operations. The ways of obtaining $l$ at the same cost are as follows:\\

	$\text{\textbf{Combination 1.}}\\
	l_1 = x_1 \oplus x_3 \text{~,~}l = l_0 \oplus l_1 \\
	\textsf{Op\_selec} = \{x_0, x_1, x_2, x_3,\ x_0 \oplus x_2,\ x_1 \oplus x_3,\ x_0 \oplus x_1 \oplus x_2 \oplus x_3 \}\\	
	\text{\textbf{~~~Combination 2.}}\\
	l_2 = l_0 \oplus x_1  \text{~,~}l = l_2 \oplus x_3  \\
	\textsf{Op\_selec} = \{x_0, x_1, x_2, x_3,\ x_0 \oplus x_2,\ x_0 \oplus x_2 \oplus x_1,\ x_0 \oplus x_1 \oplus x_2 \oplus x_3\}\\
	\text{\textbf{~~~Combination 3.}}\\
	l_3 = l_0 \oplus x_3 \text{~,~}l = l_3 \oplus x_1\\
	\textsf{Op\_selec} = \{x_0, x_1, x_2, x_3,\ x_0 \oplus x_2,\ x_0 \oplus x_2 \oplus x_3,\ x_0 \oplus x_1 \oplus x_2 \oplus x_3\}$
	
\end{example}
To guarantee optimality, we would have to test all the possibilities, but for the same reasons as above, we will only keep the first solution found each time.

\paragraph{Impact of the order of treatment of linear expressions}
The order in which we process the $l_i$ impacts the results. But note that the number of $l_i$ is large.

\begin{example}
	Let the set obtained at the end of first step be: 
	\begin{align*}
		q_0 &= x_0x_1 \oplus x_1x_2 \oplus x_1x_3 = x_1(x_0 \oplus x_2 \oplus x_3), \\
		q_1 &= x_1x_3 \oplus x_3x_4 \oplus x_1x_4 = (x_1\oplus x_3)(x_3 \oplus x_4) \oplus x_3, \\
		q_2 &= x_0x_3 \oplus x_1x_3 \oplus x_0x_4 \oplus x_1x_4  = (x_0\oplus x_1)(x_3 \oplus x_4).
	\end{align*}
	Let the output bits we want to implement be:
	\begin{align*}
		y_0 &= x_0x_1 \oplus x_1x_2 \oplus x_1x_4 \oplus x_3x_4 \oplus x_0 \oplus x_1 \oplus x_2\\
		&= q_0 \oplus q_1 \oplus x_0 \oplus x_1 \oplus x_2, \\
		y_1 &= x_0x_1 \oplus x_0x_3 \oplus x_1x_2 \oplus x_1x_3 \oplus x_0x_4 \oplus x_1 \oplus x_3\\
		&= q_0 \oplus q_1 \oplus q_2 \oplus x_1 \oplus x_4.
	\end{align*}
	$L$ will contain: 
	\begin{align*} 
		&l_0=x_0 \oplus x_2 \oplus x_3 \text{~(from $q_0$)~}\\
		&l_1=x_1 \oplus x_3 \text{~(from $q_1$)~} \\
		&l_2=x_3 \oplus x_4 \text{~(from $q_1$)~}
		&l_3=x_0 \oplus x_1 \text{~(from $q_2$)~} \\ 
		&l_4=x_0 \oplus x_1 \oplus x_2 \text{~(from $y_0$)~} \\
		&l_5=x_1 \oplus x_4 \text{~(from $y_1$)~}
	\end{align*}
Note that we did not add $x_3$ (from $q_1$) in $L$ because it is a single monomial.
\end{example}

This example is on a 2-bit S-box. On 8 bits, we observe that $L$ contains around 25 elements ($25! > 2^{83}$). We can handle separately the $l_i$ involving only 2 $x_i$ as they are necessarily obtained through a single XOR: we add these $l_i$ to \textsf{Op\_selec} directly. For all other $l_i$, we must consider order of treatment, but there are so many of them that we cannot test all permutations and restrict to a fixed number of random permutations.

\begin{algorithm}[]
	\caption{Optimisation of the linear part.}
	\label{Algo-lin}
	\SetKwInOut{Input}{Input}\SetKwInOut{Output}{Output}
	\Input{
		\begin{itemize}
			\item $L=\{l_0,....,l_m\}$: set of linear expressions to implement, induced by Algo.~\ref{Algo-non-lin},
			\item \textsf{Op\_selec}: set of selected operations \textsf{Op\_selec} initialised with $\{x_0, ... , x_{n-1}$ and expressions like $x_i \oplus x_j\}$, induced by Algo.~\ref{Algo-non-lin},
			\item $\textsf{nb\_xor\_max}$: integer, bound on the number of XOR gates given by user.
		\end{itemize}
	}
	\Output{
		\begin{itemize}
			\item A sequence of instructions to generate the set of linear expressions.
		\end{itemize}
	}

	\For {Some random permutations of $\{l_0,....,l_m\}$}
	{
		$\textsf{nb\_xor} \leftarrow \textsf{nb\_xor\_max} $ \\
		\For {$i$ from $0$ to $m$} 
		{\tcp{Browse all the linear expressions}
			\For {$j$ from $0$ to $\textsf{nb\_xor}$}
			{\tcp{Add as few XOR operations as possible}
				$S = \textsf{test\_xor\_lin} (y_i,j)$ \tcp{Test if there are linear combinations of $j$ elements equal to $l_i$}
				\If {$ S = \emptyset$}
				{
					\textbf{continue}
				}
				\Else { 
					$\textsf{nb\_xor} \leftarrow \textsf{nb\_xor} - j$ \tcp{Update the number of available operations}
					$ \textsf{Op\_selec} \leftarrow \textsf{Op\_selec}\ \cup\ S$ \tcp{Update all selected operations}
					\textbf{goto (3)} \tcp{Go back to line 3, treat the next expression}
				}
			}
			\textbf{break} \tcp{If no solutions are found, move to next permutation}
		}
		\Return \textsf{Op\_selec}
	}
\end{algorithm}

\subsection{C++ implementation of the algorithm} \label{sec:Cimplem}

\subsubsection{Implementation choices}

The code is written in C++ language for fine-tuning of memory and time.

\paragraph{Representation of polynomials}
We use 2 different representations of multivariate polynomials. The first allows to represent full algebraic normal forms (ANF), the second is optimised for memory space: it represents polynomials of fixed-degree (1 or 2).

In the first representation, a multivariate polynomial $p$ is a vector of $2^n$ Boolean values, stored as an array $p$ of \textsf{uint32\_t}.
The value of bit $i$ of $p[j]$ corresponds to the presence or absence of monomial $x^{32j+i}$, where $x^u = \prod_i(x_i^{u_i})$, with $u_i$ the bits of $u$. For example, $5 = (101)_2$, so the monomial $x^5$ represents $(x_0)^1.(x_1)^0.(x_2)^1 = x_0x_2$.

The ANF of an $n$-bit function $S$ is made of $n$ lines, each of them corresponding to the ANF of the coordinate function $S_i$ (\textit{i.e.}~the polynomial of the $i$-th output bit).

\begin{example}[ANF representation]
	Over 3 bits, there are $2^3 = 8$ monomials. 
	The polynomial $1 \oplus x_0 \oplus x_1x_2 \oplus x_0x_1x_2$ is encoded as:
	\vspace{0.2cm}
	\begin{center}
		\begin{tabular}{|c|c|c|c|c|c|c|c|}
			\hline
			$x^0$ & $x^1$ & $x^2$ & $x^3$ & $x^4$ & $x^5$ & $x^6$ & $x^7$\\
			\hline
			$1$ & $x_0$ & $x_1$ & $x_0x_1$ & $x_2$ & $x_0x_2$ & $x_1x_2$ & $x_0x_1x_2$\\
			\hline
			1 & 1 & 0 & 0 & 0 & 0 & 1 & 1 \\
			\hline
		\end{tabular}
	\end{center}
	
\end{example}
\vspace{0.2cm}
The second representation is for polynomials of degree 1 or 2. Quadratic polynomials are encoded as positive integers of $32$ or $64$ bits. Each bit of the integer encodes presence of absence of each \textit{quadratic} monomial. The monomials are lexicographically ordered.

\begin{example}[Degree-2 representation]
	Over 5 bits, there are $\binom{5}{2} = 10$ quadratic monomials, hence, a polynomial can be stored as 10 bits, each encoding the presence or absence of a monomial.
	Here is an encoding of polynomial $x_0x_4 \oplus x_1x_3 \oplus x_1x_4 \oplus x_2x_4$:
	\vspace{0.2cm}
	\begin{center}
		\begin{tabular}{|c|c|c|c|c|c|c|c|c|c|}
			\hline
			$x_0x_1$ & $x_0x_2$ & $x_0x_3$ & $x_0x_4$ & $x_1x_2$ & $x_1x_3$ & $x_1x_4$ & $x_2x_3$ & $x_2x_4$ & $x_3x_4$ \\
			\hline
			0 & 0 & 0 & 1 & 0 & 1 & 1 & 0 & 1 & 0 \\
			\hline
		\end{tabular}
	\end{center}
	\vspace{0.2cm}
	This polynomial is represented by the integer $(0001011010)_2 = (90)_{10}$.
\end{example}

Encoding of linear polynomials is similar, each bit corresponding to a monomial $x_i$.

\begin{example}[Degree-1 representation]
	Over 5 bits, $x_0 \oplus x_2 \oplus x_3$ has encoding:
	\vspace{0.2cm}
	\begin{center}
	\begin{tabular}{|c|c|c|c|c|}
		\hline
		$x_0$ & $x_1$ & $x_2$ & $x_3$ & $x_4$ \\
		\hline
		1 & 0 & 1 & 1 & 0 \\
		\hline
	\end{tabular}
\end{center}
	\vspace{0.2cm}	
	It is therefore represented by the integer $(10110)_2 = (22)_{10}$.
\end{example}

\paragraph{Precomputation}
In order to optimise the time complexity, we decided to precompute some data common to all executions (for a given S-box input size):
\begin{itemize}
	\item \textsf{Set\_op} containing all the operations that can be obtained in 1 AND,
	\item \textsf{Map\_xor} containing all XOR of pairs of elements of \textsf{Set\_op}. It allows to test if an expression is equal to the XOR of a pair of elements of \textsf{Set\_op}, and gives the pair.
\end{itemize}

\begin{example}[\textsf{Map\_xor}]
On 7 bits, the operation encoded by $(393237)_{10}$ represents $x_0x_3 \oplus x_0x_4 \oplus x_3x_5 \oplus x_4x_5 \oplus x_5x_6$. It is an element of \textsf{Map\_xor}, obtained by the pairs:

\begin{table}[H]
\begin{center}
	\begin{tabular}{|c|c|c|c|}
		\hline
		First operation & Encoding & Second operation & Encoding \\
		\hline
		$x_5x_6$ & 1 & $(x_0 \oplus x_3)(x_4 \oplus x_5)$ & 393236 \\
		\hline
		$x_6(x_3 \oplus x_4 \oplus x_5)$ & 11 & $(x_0 \oplus x_5 \oplus x_6)(x_3 \oplus x_4)$ & 393246 \\
		\hline
		$x_5(x_3 \oplus x_4 \oplus x_6)$ & 21 & $x_0(x_3 \oplus x_4)$ & 393216 \\
		\hline
		$(x_3 \oplus x_4 \oplus x_5)(x_5 \oplus x_6)$ & 31 & $(x_0 \oplus x_6)(x_3 \oplus x_4)$ & 393226 \\
		\hline
		$x_0x_6$ & 32768 & $(x_0 \oplus x_5)(x_3 \oplus x_4 \oplus x_6)$ & 426005 \\
		\hline
		$x_6(x_0 \oplus x_3 \oplus x_4)$ & 32778 & $(x_0 \oplus x_5 \oplus x_6) \times $ & 426015 \\
		&& $(x_3 \oplus x_4 \oplus x_6)$ & \\
		\hline
		$x_5(x_0 \oplus x_6)$ & 65537 & $(x_0 \oplus x_5)(x_3 \oplus x_4 \oplus x_5)$ & 458772 \\
		\hline
		$x_5(x_0 \oplus x_3 \oplus x_4 \oplus x_6)$& 65557 & $x_0(x_3 \oplus x_4 \oplus x_5)$ & 458752 \\
		\hline
		$x_0(x_5 \oplus x_6)$ & 98304  & $(x_0 \oplus x_5) \times$ & 491541 \\
		&& $(x_3 \oplus x_4 \oplus x_5 \oplus x_6)$ & \\
		\hline
		$(x_5 \oplus x_6)(x_0 \oplus x_3 \oplus x_4)$ & 98334 & $(x_0 \oplus x_6) \times $ & 491531 \\
		&& $(x_3 \oplus x_4 \oplus x_5 \oplus x_6)$ & \\
		\hline
	\end{tabular} 
\end{center}
\vspace{0.2cm}
The line of \textsf{Map\_xor} associated with 393237 will contain: \\
$\{1,393236\}\ \{11,393246\}\ \{21,393216\}\  \{31,393226\}\ \{32768,426005\}\ $ \\
$\{32778,426015\} \{65537,458772\}\{65557,458752\}\ \{98304,491541\}\ $ \\ $\{98334,491531\}$
\end{table}

\label{Example of map_xor}
\end{example}
Storing this map helps to speed up the $\textsf{test\_xor}$ function. Indeed, to test if an operation $x$ can be obtained by adding 2 operations of \textsf{Set\_op}, instead of Algo.~\ref{Without_map}, we can use Algo.~\ref{with_map}.

\begin{algorithm}[]
	\caption{Example of $\textsf{test\_xor}$ without using \textsf{Map\_xor}.}
	\label{Without_map}
	\For {$i$ from 0 to \textsf{Set\_op\_size}}
	{
		$\textsf{elem} \leftarrow x \oplus \textsf{Set\_op}[i]$ \\
		$\textsf{index} \leftarrow \textsf{Set\_op}.\textsf{find}(\textsf{elem})$ \\
		$\textsf{solution} \leftarrow \textsf{solution}\ \bigcup\ \{ \textsf{Set\_op}[i],\ \textsf{Set\_op}[\textsf{index}]\}$ \\
	}
	\Return $\textsf{solution}$
\end{algorithm} 

\begin{algorithm}[]
	\caption{Example of $\textsf{test\_xor}$ using \textsf{Map\_xor}.}
	\label{with_map}
	$\textsf{index} \leftarrow \textsf{Map\_xor.\textsf{find}(x})$ \\
	\For {$i$ from 0 to \textsf{Map\_xor}[\textsf{index}].size}
	{
		$\textsf{solution} \leftarrow \textsf{solution}\ \bigcup\ $ \\
		$ \{ \textsf{Map\_xor}[\textsf{index}][i][0], \textsf{Map\_xor}[\textsf{index}][i][1]\}$ \\
	}
	\Return $\textsf{solution}$
\end{algorithm}

Sadly we cannot do the same for storing all the XOR of 3 elements of \textsf{Set\_op}, as the number of triples is far too large.
Details on the sizes of \textsf{Set\_op} and \textsf{Map\_xor} are given in Table~\ref{table:set_op_map_xor}, together with the size of a hypothetical map of triples of \textsf{Set\_op} (called \textsf{Triples}). No value for \textsf{Triples} is given when \textsf{Set\_op} and \textsf{Map\_xor} cover all quadratic functions.

\begin{table}[H]\setlength\tabcolsep{.5em}
	\begin{center}
	\begin{tabular}{cccc}
		\hline
		\multirow{2}*{Bits} & \multirow{2}*{$\#$\textsf{Set\_op}} & \multicolumn{2}{c}{$\#$\textsf{Map\_xor}} \\
		&& Number of keys & Size (bytes) \\
		\hline
		4 & 35 & 35 & 1 KB \\
		\hline
		5 & 155 & 868 & 17 KB \\
		\hline
		6 & 651 & 18 228 & 356 KB  \\
		\hline
		7 & 2667 & 330 708 & 6 MB \\
		\hline
		8 & 10 795 & 5 622 036 & 107 MB \\
		\hline
		9 & 43 435 & 92 672 916 & 2 GB  \\
		\hline
		10 & 172 846 & $>$1 505 005 887 & $>$ 28 GB \\
		\hline
		11 & 692 775 &&\\
		\hline
	\end{tabular}
	\end{center}
	\vspace{0.2cm}
	\begin{center}
	\begin{tabular}{ccc}
		\hline
		\multirow{2}*{Bits} &   
		\multicolumn{2}{c}{$\#$\textsf{Triples}} \\
		& Number of keys & Size (bytes) \\
		
		\hline
		4 & - & - \\
		\hline
		5 & - & - \\
		\hline
		6 & 13 328 & 43 MB \\
		\hline
		7 & 1 761 110 & 6 GB\\
		\hline
		8 && \\
		\hline
		9 && \\
		\hline
		10 &&\\
		\hline
		11 &&\\
		\hline
	\end{tabular}  
	\end{center}
	\caption{Details of precomputed data sizes.}
	\label{table:set_op_map_xor}
\end{table}

\subsubsection{Data structures}

\paragraph{\textsf{Set\_op} and \textsf{Map\_xor} during run}
\textsf{Set\_op} and \textsf{Map\_xor} are large, but precomputed and fixed. As such, during a run of the tool, they are stored as static arrays. These are the most important data structures: they are the largest and the most accessed. Being able to precompute and store them statically is our most powerful data-structure optimisation.

\paragraph{Formalising the data-structure problem} Static arrays are perfect to optimise most of the data during run and do not require much understanding of the context, but this data must first be precomputed. During precomputation, we cannot predict the contents or even the sizes of \textsf{Set\_op} and \textsf{Map\_xor}, hence arrays are out of the question. We must use other data structures, and \emph{this is the point that limits our tool to S-boxes of less than 10 bits}.
We will explain the problem and possible solutions in the next paragraphs, focusing on \textsf{Map\_xor} which is the largest data structure to be computed.

\paragraph{Precomputation cases}
We consider 4 cases. Case 1 is the ideal case: building the sets and maps for \textit{all} $n$-bit polynomials (to implement any $n$-bit S-box). Case 2 is building sets and maps for all \textit{quadratic} $n$-bit polynomials (without linear part), Case 3 is for quadratic $n$-bit polynomials \textit{obtainable in 1 AND} and Case 4 is for all quadratic \textit{monomials}.

In practice, we use case 3 during run, and a variant of it during precomputation, but the other options are useful to understand why this choice matters.

\paragraph{What do we want to store?} What we want is to know which combination of polynomials can lead to which polynomials. This could be represented as a graph, in which nodes are polynomials and edges are the relation ``node $x$ can lead to node $y$ through a XOR''. However we already have a problem with this representation: a binary gate (like XOR) is a relation between 3 variables (2 inputs and 1 output), not 2 like an edge in our graph. There is a (dirty) way of storing this relation in a graph: duplicating all edges. To store that $x \oplus x' = y$, we store 2 labeled edges $(x\xrightarrow{x'} y)$ and $(x'\xrightarrow{x} y)$.

\paragraph{Storing the graph}
To get an idea of the size of the data structures in each case, let us do some basic combinatorics. Let $n$ be the input size of the S-box and $N_i$ be the number of polynomials to store for Case number $i$. Values of $N_i$ are summarised in Table~\ref{table:ni}.

\begin{center}
\begin{table}[H]\setlength\tabcolsep{.8em}\centering
	\begin{tabular}{|l|cccc|}
		\hline
		& $N_1$ & $N_2$ & $N_3$ & $N_4$ \\
		\hline
		&&& see Table~\ref{table:set_op_map_xor}, & \\
		Formula & $2^{2^n}$ & $2^{n(n-1)/2}$ & ``$\#$\textsf{Map\_xor}'' & $n(n-1)/2$\\
		$n=7$ & $2^{128}$ & $2^{21}$ & 330 708 & $21$\\
		$n=10$ & $2^{1024}$ & $2^{45}$ & $>$1 505 005 887 & $45$\\
		\hline
	\end{tabular}
	\caption{Number of polynomials.}
	\label{table:ni}
\end{table}
\end{center}

As we are interested in actual data size in bits, let us add that storing 1 polynomial of case $i$ takes $log_2(N_i)$ bits (in practice, this is rounded up to a multiple of 32).

\paragraph{Storage as adjacency matrix} A classical way to store a graph is as an adjacency matrix. With the above representation of the graph with duplicated edges, a cell of the matrix corresponds to an edge and this cell stores the label of the edge (which is an input value, hence a polynomial/monomial).

Then there are $N_i^2$ cells in the matrix, and each cell takes $log_2(N_i)$ bits, hence the whole matrix takes $log_2(N_i)\times N_i^2$ bits.

As we can see in Table~\ref{table:graph_storage}, only the minimal precomputation of case 4 is doable on 10 bits. Case 2 is doable up to $n=7$, in which case the matrix can be stored with about $2^{47}$ bits in memory, stored as \texttt{uint32\_t}, which is about 16TB.

\paragraph{Storage as adjacency vectors} The other classical way to store a graph is using several adjacency vectors: for each node of the graph (\textit{i.e.} each polynomial), we store a linked list of its successors in the graph. Formally, consider $x$ a polynomial, we build a linked list of its successors $y$ (\textit{i.e.} the polynomials that $x$ can reach in one XOR gate). Then each vector cell stores a pair $\{y,next\}$ , where $next$ serves to implement the linked list (it is the address of the next element in the list).

Each cell then takes $S=log_2(N_i)+64$ bits in memory, assuming a 64-bit address size. Note that in practice, $log_2(N_i)=64$ for large $n$ (or 32 for small $n$), thus $S\simeq2^7$. In particular, note that storing $next$ doubles the memory consumption.

Storing with vectors only makes sense if the adjacency matrix is sufficiently sparse. The question is thus: On average, how many polynomials $y$ can a polynomial $x$ reach through one XOR gate?

Interestingly, in practice for case 3 (quadratic polynomials obtainable in 1 AND), we observe that the number of $y$.s reachable by any $x$ is constant, equal to 10, for any $n$. We did not push the theory further to prove this, as it is only a detail in our work, but we conjecture that this number is always equal to 10. We expect that this could be proved using code theory, as the problem can be translated in terms of quadratic Reed-Solomon code, punctured over all linear polynomials.

With this conjecture, we will add this constant $M=10$ to our study, and deduce that the matrix is very sparse. Note that it is only a good estimate for case 3 and limits our tool to 9-bit S-boxes.

The summary of memory consumption can be found in Table~\ref{table:graph_storage}.
Overall, we use far less memory than with matrices and lose little time in data access. For $n=10$, it is still doable: $2^{41}\simeq$ 250GB for quadratic polynomials obtainable in 1 AND.

\begin{center}
\begin{table}[H]\setlength\tabcolsep{.7em}\centering
	\begin{tabular}{lcccc|cccc}
		\hline
		& \multicolumn{4}{c}{Matrix} & \multicolumn{4}{c}{Vectors}\\
		& $N_1$ & $N_2$ & $N_3$ & $N_4$ & $N_1$ & $N_2$ & $N_3$ & $N_4$ \\
		\hline
		$n=7$ & $2^{263}$ & $2^{47}$ & $2^{41}$ & $2^{11}$ & $2^{138}$ & $2^{31}$ & $2^{28}$ & $2^{15}$\\
		$n=10$ & $2^{2058}$ & $2^{96}$ & $2^{76}$ & $2^{14}$ & $2^{1034}$ & $2^{55}$ & $2^{41}$ & $2^{17}$\\
		\hline
	\end{tabular}
	\caption{Graph storage sizes (orders of magnitude).}
	\label{table:graph_storage}
\end{table}
\end{center}

Note that actually, we know that vectors are of size $M$ cells, with $M=10$ constant, thus we instantiate the vectors with arrays of $M$ cells (and we keep the arrays sorted). This way, access time is reduced: with vectors, we would have on average $M/2$ cell accesses to find an element, while with arrays we end up with about $log_2(M)/2$ cell accesses. Even better, with vectors memory is non-contiguous, so on average every one of the $M/2$ cell accesses are cache misses, while arrays are contiguous in memory, thus we make $log_2(M)/2$ accesses where only the first one is a cache miss. Overall, element access in our adjacency vectors takes only 1 RAM access (and a few cache accesses): similar to the adjacency matrix case but without wasting memory.

Also, with arrays rather than vectors, each cell contains only $\{y\}$ and not the adress of the next vector element, which divides memory size by a factor roughly 2.

\paragraph{From theory to practice} However, during precomputation we cannot use case 3 directly, as we do not know the contents or size of the maps in advance (we have not found a theoretical estimate of the sizes yet). We could use case 2, which would be doable up to $n=9$ (with about 8TB).

An alternative is to not use static arrays to store the data, but dynamic data structures in which we insert the elements when first precomputing them. In this case, we require (1) uniqueness of the elements and (2) sorted elements for efficient ``find'' and ``insert'' procedures. Overall, we need \textit{ordered sets} (or ordered maps, which are ordered sets of pairs).

There is however a problem with [ordered] sets in practice: we could not find any efficient implementation for them. Indeed, the standard C++ \texttt{set} is highly inefficient in memory: it is stored as a binary search tree, hence each node contains its 32- or 64-bit polynomial, plus two 64-bit addresses for its left and right children. This results in 160 bits to store a 32-bit polynomial, or 192 bits to store a 64-bit polynomial, \emph{a loss of memory by a factor between 3 and 5}.

The only alternative would be to rewrite an efficient implementation of ordered sets ourselves, however this is a big work in itself, and not the main topic of this article. For this reason, we stick with C++ \texttt{set}, but we lose a factor 3 in memory, which is highly limiting (we jump from 250GB for $n=10$ to 750GB, as a lower bound).

\paragraph{Summing up} We use case 3 to store \textsf{Set\_op} and \textsf{Map\_xor} during run, stored as static arrays with overall memory complexity of $N_3\times M\times S$ bits. During precomputation however, we are limited by the necessary use of ordered sets which increase memory usage by a factor at least 3.

\subsubsection{Parallelisation}

Several parts of the code were parallelised using \texttt{openmp}.

The first idea was to parallelise when all the permutations of the output bits are run through, mainly for implementation reasons. The way each permutation is processed is identical, so this solution is simple to implement and allows less interaction between threads. This is the option chosen for the linear part. The parallelisation then takes place at line 1 of Algorithm~\ref{Algo-lin}.
For the non-linear part, we use a different setting. The reason is that, in cases where being exhaustive is costly, it makes more sense to try to target distant parts of the search space to cover more potentially different implementations. The permutations are run through after the initialisation of \textsf{Op\_selec}. Therefore, parallelising at this point and starting the search with different sets allows us to reach different local optima which will allow more variety on final solutions. This configuration seems to have a greater impact than the other version on obtaining a solution. The parallelisation then takes place at line 2 of Algorithm \ref{Algo-non-lin}.

\begin{example}
Let $y_i = x_0x_3 \oplus x_0x_4 \oplus x_3x_5 \oplus x_4x_5 \oplus x_5x_6 = (393237)_{10}$ an output bit to be implemented. Each thread will begin their search with \textsf{Op\_selec} initialised with one of the couples detailed in Example \ref{Example of map_xor}.
\end{example}

\subsubsection{Search-space limitation for non-exhaustive search}

\paragraph{Limiting factors}

From 8 bits upwards, there are quadratic Boolean functions that cannot be obtained by XORing only three 1-AND-gate Boolean functions. The search must therefore be initialised using quadruples, but the number of distinct quadruples needed to obtain a given operation is around 350,000. This is too much to be exhausted, so exhaustive search is no longer possible.

Also, remember that we need to exhaust all the possible permutations of the output bits because the order of processing matters. However the time required to test a permutation increases with the number of bits and the size of the tuple initialising the search. Details of average processing times for one permutation are given in Table~\ref{table3}.
As we can see, it also becomes impossible to exhaust all permutations of the output bits from 8 bits upwards.

\begin{table}[H]\setlength\tabcolsep{.5em}
	\begin{center}
		\begin{tabular}{ccc}
			\hline
			& Average processing & Average processing \\
			Number of bits & time for one & time for one \\
			& permutation initialised & permutation initialised \\
			& with a triple &  with a quadruple \\
			\hline
			7 & 1 sec & - \\
			\hline
			8 & ~10 hours & ~ 18 hours \\
			\hline
			9 & ~30 hours  &  $>$ 60 hours\\
			\hline
		\end{tabular} 
		\caption{Details of average processing times for a permutation.}
		\label{table3}
	\end{center}
\end{table}

\paragraph{Search-space bound}

Since exhaustive search is impossible from 8 bits upwards, we instead consider a random limited part of the search space.

We have chosen to use only a fixed number \textsf{NB\_QUAD\_MAX} of quadruples drawn at random. The default value of \textsf{NB\_QUAD\_MAX} is 20,000.

We only consider a fixed number \textsf{NB\_PERM\_MAX} of output bits permutations, drawn at random. \textsf{NB\_PERM\_MAX} is defined differently for each configuration by a macro in the program header. As our experimentation setup only allows us to treat S-boxes up to 9 bits, we have not defined any limits for cases where the size is greater than 9 bits. The values given in Table~\ref{table4} are the default values.
\begin{table}[H]\setlength\tabcolsep{.5em}
	\begin{center}
		\begin{tabular}{cccc}
			\hline
			Number of bits && \textsf{NB\_PERM\_MAX} & \\
			& for a couple & for a triple & for a quadruple \\
			\hline
			8 & 50 & 10 & 5 \\
			\hline
			9 & 50 & 5 & 2  \\
			\hline
		\end{tabular} 
	\end{center}
	\caption{Default value for \textsf{NB\_PERM\_MAX}.}
	\label{table4}
\end{table}

\paragraph{About suboptimal result} 

From 8 bits upwards, we can no longer guarantee the optimality of the results, as the search is no longer exhaustive. The implementations given in Section~\ref{sec:results} for 8 and 9 bits are the best we obtained, but better ones may exist.

\subsubsection{How to use the tool}

\begin{table*}
	\begin{center}
		\begin{tabular}{cccc|cccc|ccc}
			\hline
			&&&&\multicolumn{4}{c}{Implementation(AND - XOR)} & \multicolumn{3}{c}{Timings (in \textbf{seconds})} \\
			S-box & $\delta$ & $\mathcal{L}$ & Bij & \cite{DBLP:journals/tosc/BilginMDLS20} &  \cite{DBLP:journals/iacr/ZhangH23} & \cite{DBLP:journals/tcasI/FengWZPZ24}& Ours & \cite{DBLP:journals/iacr/ZhangH23}&
			\cite{DBLP:journals/tcasI/FengWZPZ24} & Ours \\
			\hline
			$\chi_5$ & 8 & 16 & Yes & - & 5 - & 5 - 5 & 5 - 10 & 1.2 & 19.81 & 0.05 \\
			\hline
			ASCON & 8 & 16 & Yes &  -  & 5 -  & 5 - & 5 - 15 & 1.4 & 26.48 & 0.1 \\
			\hline
			SYCON & 8 & 16 & Yes &  -  & -  & 5 - 17 & 5 - 15 &  & 26.72 & 0.1 \\
			\hline
			FIDES & 2 & 8 & Yes & 7 - 29 & 7 - & 7 - 27 & 7 - 20 & 3.6 & 140.31 & 0.2 \\
			\hline
			$X^3$ & 2 & 8 & Yes & 7 - 29 & - & - & 7 - 19 &  &  & $<$ 1 \\
			\hline
			$X^5$ & 2 & 8 & Yes & 7 - 26 & - & - & 7 - 21 &  &  & $<$ 1 \\
			\hline
			S\_51 & 4 & 16 & Yes & - & - & - & 5 - 20 &  &  & $<$ 1 \\
			\hline
			S\_52 & 4 & 16 & Yes & - & - & - & 5 - 24 &  &  & $<$ 1 \\
			\hline
			S\_53 & 4 & 16 & Yes & - & - & - & 5 - 24 &  &  & $<$ 1 \\
			\hline
			S\_54 & 8 & 16 & Yes & - & - & - & 6 - 20 &  &  & $<$ 1 \\
			\hline
		\end{tabular}
		\caption{Results for 5-bit S-boxes.} \label{results5}
	\end{center}
\end{table*}

Our tool takes as inputs the look-up table of a given S-box and a maximum runtime and returns a C-program implementing the S-box if a solution is found. 

Several other arguments can be added:

\begin{list}{}{}
	\item{\texttt{--solmax:} the number of desired solutions,}
	\item{\texttt{--andmax:} an upper bound on the number of AND gates,}
	\item{\texttt{--precomputation\_files 0:} if we do not want to use the precomputation files.}
\end{list}

By default, the program will use the precomputation files and return all the optimal solutions it can find. The precomputation files are not given with the code, due to their large size, but a \texttt{Makefile} is provided to compute them. The size of these files is given in Table ~\ref{table5}. Note that the RAM size of the computer must be larger than these sizes to run the tool.

\begin{table}[H]
	\begin{minipage}{.4\linewidth}
	\begin{center}
		\begin{tabular}{|c|c|}
			\hline
			Number of bits & Size   \\
			\hline
			4 & 20 KB \\
			\hline
			5 & 88 KB \\
			\hline
			6 & 2216 KB \\
			\hline
		\end{tabular} 
		\end{center}
	\end{minipage} \hfill
\begin{minipage}{.4\linewidth}
	\begin{center}
		\begin{tabular}{|c|c|}
			\hline
			Number of bits & Size  \\
			\hline
			7 & 51 MB \\
			\hline
			8 & 1.1 GB\\
			\hline
			9 & 22 GB \\
			\hline
		\end{tabular} 
	\end{center}
\end{minipage}
	\caption{Size of the precomputation files.}
	\label{table5}
\end{table}

\begin{example}
Let $S$ be an S-box defined by the following look-up-table (in little endian): \\
\begin{center}
\begin{tabular}{|c|c|c|c|c|c|c|c|c|c|c|c|c|c|c|c|}
	\hline
	0&1&2&3&4&5&6&7&8&9&A&B&C&D&E&F \\
	\hline
	0&1&4& 5& 3& 2& 7& 6& C& D& 8& 9& F& E& B& A \\
	\hline
\end{tabular}
\end{center}
\vspace{0.2cm}
If we want to obtain at most 3 implementations of S using 4 AND gates, not wait more than 10 minutes and not use the precomputation files, the command line to execute the program will be: \\
\texttt{./implem -\--solmax\ 3 -\--andmax\ 4 -\--lut\ 0,1,4,5,3,2,7,6,12,13,8,9,15,14,11,10 -\--timemax\ 10 -\--precomputed\_files\ 0}
\end{example}

\section{Results}\label{sec:results}
\noindent

We give a few experimental results that we obtained for up to 9-bit S-boxes. We chose S-boxes covering different levels of security. Notably, we used power functions and XORs of power functions to get good cryptographic properties, as they are well studied. We compare our implementations with the ones given in~\cite{DBLP:journals/tosc/BilginMDLS20}, \cite{DBLP:journals/iacr/ZhangH23} and \cite{DBLP:journals/tcasI/FengWZPZ24}. We also compared them with the implementations obtained using the ABC \footnote{\url{https://github.com/berkeley-abc/abc}} tool in order to compare with a non-cryptographic approach. Execution times are reported when available. Note that the tool runtime in itself is a secondary objective (as they serve in algorithms used billions of times per second). Runtime is mostly a useful indicator of whether the approach scales reasonably, since execution time is one of the main limiting factors for large input sizes.
\paragraph{Experimentation setup}

The tool was used on 2 different types of machine:
\begin{itemize}
	\item Config 1: 64-bit x86 with an Intel core i7 at 2.3GHz, with 12 cores and 32GB of RAM.
	\item Config 2: 64-bit x86 with an AMD EPYC processor at 2.8GHz, with 64 cores and 128GB of RAM.
\end{itemize}

We chose our 2 configurations to be representative of what anyone can do with a common personnal computer (Config.~1) and what is doable with more expensive means (Config.~2).

It should be noted that our hardware environnement may differ from the conditions used in other papers. In \cite{DBLP:journals/tosc/BilginMDLS20}, the hardware environnement is not specified. In \cite{DBLP:journals/tcasI/FengWZPZ24}, the experiments were performed on an Intel Xeon CPU at 2.1GHz, with 8 cores. In \cite{DBLP:journals/iacr/ZhangH23}, the experiments were performed on an Intel core i7 at 3.4GHz and an AMD 3990X CPU at 2.9GHz and unspecified number of cores.

\paragraph{Memory used}
In config 2 (128 GB RAM), up to 8 bits, the RAM used is less than 1\%. For 9 bits, the program requires around 14\% of the available RAM.

\paragraph{Results tables}
The Tables \ref{results5}, \ref{results6}, \ref{results7}, \ref{results8}, \ref{results9} detail results for several S-boxes, specifying their differential uniformity ($\delta$), linearity ($\mathcal{L}$), whether or not they are bijective and the tool timing. For sizes 5 and 6, as the timings are very similar between the two configurations, only those for configuration 1 are given.  AND depth is not specified as it is always 1.

All the corresponding look-up-tables and implementations can be found as Supplementary Material. The $\chi_i$ S-boxes for $5\leq i\leq 9$ bits corresponds to the transformation used for instance in Keccak and defined as in~\cite{DBLP:conf/space/DaemenMM21}.

Note that for $\chi_i$, our results use double the number of XOR compared to~\cite{DBLP:conf/space/DaemenMM21}. This actually comes from the choice of linear expressions in which we choose not to care about NOT gates (see Footnote~\ref{footnoteNOT}).

\subsection{Over 5 bits (Table \ref{results5})}
S-boxes S\_51 to S\_54 are randomly generated.
Power maps use the field polynomial $X^5+X^2+1$.

\subsection{Over 6 bits (Table \ref{results6})}

S-boxes Q2256 to Q2263 are representatives of the 8 extended-affine-equivalence classes of permutations with the best differential uniformity and linearity~\cite{DBLP:journals/iacr/MeyerB18}.
S-boxes BL\_ECCZ\_1 to BL\_ECCZ\_14 are representatives of the 13 CCZ-equivalence classes of quadratic APN functions~\cite{DBLP:journals/dcc/BrinkmannL08}.
Power maps use the field polynomial $X^6+X^4+X^2+X+1$.
S-boxes S\_61 and S\_62 are randomly generated.

\begin{table}
	\begin{center}
		\adjustbox{valign=t,max width=.95\columnwidth}{
		\begin{tabular}{cccc|ccc|c}
			\hline
			&&&&\multicolumn{3}{c}{Implementation(AND - XOR)} & \multicolumn{1}{c}{Our timings} \\
			S-box & $\delta$ & $\mathcal{L}$ & Bij & \cite{DBLP:journals/tosc/BilginMDLS20} & \cite{DBLP:journals/tcasI/FengWZPZ24} & Ours & (in \textbf{minutes}) \\
			\hline
			$\chi_6$ & 16 & 32 & No & - & 6 - & 6 - 12 & $<$ 1 \\
			\hline
			Q2256 & 4 & 16 & Yes & 8 - 35 & - & 8 - 26 & $<$ 1 \\
			\hline
			Q2257 & 4 & 16 & Yes & 8 - 41 & - & 8 - 27 & $<$ 1 \\
			\hline
			Q2258 & 4 & 16 & Yes & 8 - 38 & - & 8 - 25 & $<$ 1 \\
			\hline
			Q2259 & 4 & 16 & Yes &  -  & - & 8 - 28 & $<$ 1 \\
			\hline
			Q2260 & 4 & 16 & Yes & 8 - 40 & - & 8 - 29 & $<$ 1 \\
			\hline
			Q2261 & 4 & 16 & Yes &  -  & - & 8 - 26 & $<$ 1 \\
			\hline
			Q2262 & 4 & 16 & Yes & -  & - & 8 - 27 & $<$ 1 \\
			\hline
			Q2263 & 4 & 16 & Yes & 8 - 38 & - & 8 - 30 & $<$ 1 \\
			\hline
			$X^3$ & 2 & 16 & No & 9 - 43 & - & 8 - 37 & $<$ 1 \\
			\hline
			$X^9$ & 8 & 64 & No & - & - & 6 - 28 & $<$ 1 \\
			\hline
			BL\_ECCZ\_1 & 2 & 16 & No & - & - & 9 - 23 & $<$ 1 \\
			\hline
			BL\_ECCZ\_8 & 2 & 16 & No & - & - & 8 - 34 & $<$ 1 \\
			\hline
			BL\_ECCZ\_11 & 2 & 16 & No & - & - & 8 - 29 & $<$ 1 \\
			\hline
			BL\_ECCZ\_14 & 2 & 16 & No & - & - & 9 - 23 &  $<$ 1 \\
			\hline
			S\_61 & 16 & 32 & No & - & - & 6 - 17 & $<$ 1 \\
			\hline
			S\_62 & 8 & 32 & No & - & - & 7 - 24 & $<$ 1 \\
			\hline
		\end{tabular} 
	}
		\caption{Results for 6-bit S-boxes.} \label{results6}
	\end{center}
\end{table}

\subsection{Over 7 bits (Table \ref{results7})}

S-boxes S\_71 to S\_76 are randomly generated.
Power maps use the field polynomial $X^7+X+1$.

\begin{table}
	\begin{center}
	\adjustbox{valign=t,max width=.95\columnwidth}{\setlength\tabcolsep{.15em}
		\begin{tabular}{cccc|ccc|cc}
			\hline
			&&&&\multicolumn{3}{c}{Implem.(AND - XOR)} & \multicolumn{2}{c}{Our timings (in \textbf{min})} \\
			S-box & $\delta$ & $\mathcal{L}$ & Bij & \cite{DBLP:journals/tosc/BilginMDLS20} & \cite{DBLP:journals/tcasI/FengWZPZ24} &  Ours &   Config 1 &  Config 2 \\
			\hline
			$\chi_7$ & 32 & 64& Yes && 7 - & 7 - 14 & $<$ 1 & $<$ 1\\
			\hline
			$X^3$ & 2 & 16 & Yes & 15 - 79 & - & 11 - 45 & & 3 \\
			\hline
			$X^5$ & 2 & 16 & Yes & - & - & 11 - 56  & 4 & 1 \\
			\hline
			$X^9$ & 2 & 16 & Yes & - & - & 11 - 53  & 3 & 2 \\
			\hline
			$X^{17}$ & 2 & 16 & Yes & - & - & 11 - 54  &  & 1 \\
			\hline
			$X^{33}$ & 2 & 16 & Yes & - & - & 11 - 58  &  & 2 \\
			\hline
			$X^{65}$ & 2 & 16 & Yes & - & - & 11 - 52  &  & 2 \\
			\hline
			$X^3 \oplus X^5$ & 4 & 32 & No & - & - & 10 - 57 &  & 68 \\
			\hline
			S\_71 & 8 & 32 & No & - & - & 10 - 51 &  & 2 \\
			\hline
			S\_72 & 8 & 64 & No & - & - & 10 - 48  & 120 & 10 \\
			\hline
			S\_73 & 4 & 32 & No & - & - & 11 - 50  &  & 2 \\
			\hline
			S\_74 & 32 & 64 & No & - & - & 7 - 30  & $<$ 1 & $<$ 1 \\
			\hline
			S\_75 & 16 & 64 & No & - & - & 8 - 31  & $<$ 1 & $<$ 1 \\
			\hline
			S\_76 & 8 & 32 & No & - & - & 10 - 43  & $<$ 1 & $<$ 1 \\
			\hline
		\end{tabular}
		}
		\caption{Results for 7-bit S-boxes.} \label{results7}
	\end{center}
\end{table}

\subsection{Over 8 bits (Table \ref{results8})}

S-box S\_A81 is a representative of one of the (approximately) 13,000 classes of quadratic APN functions for CCZ equivalence, classified in~\cite{DBLP:journals/tit/BeierleL22}.
S-boxes S\_81 to S\_83 are randomly generated. $X^3$ truncated is $X^3$ over 9 bits where we truncate the 8th bit (most significant). Power maps use the field polynomial $X^8+X^4+X^3+X^2+1$.

These results are the best we obtained, but better implementations may exist. We tried to obtain one AND less, limiting the runtime to a few days, without success.

\begin{table} 
	\adjustbox{valign=t,max width=.95\columnwidth}{
		\begin{tabular}{cccc|cc|c}
			\hline
			&&&&\multicolumn{2}{c}{Implementation} & \multicolumn{1}{c}{Our timings} \\
			S-box & $\delta$ & $\mathcal{L}$ & Bij & \multicolumn{2}{c}{(AND - XOR)} & \multicolumn{1}{c}{(in \textbf{minutes})} \\ &&&&\cite{DBLP:journals/tcasI/FengWZPZ24} &  Ours & Config 2 \\
			\hline
			$\chi_8$ & 64 & 128 & No & 8 - & 8 - 16 & $<$ 1\\
			\hline
			$X^3$ & 2 & 32 & No & - & 14 - 77 & 115 \\
			\hline
			$X^5$ & 4 & 64 & No & - & 12 - 71 & 3  \\
			\hline
			$X^9$& 2 & 32 & No & - & 14 - 75 & 900 \\
			\hline
			$X^{17}$ & 16 & 256 & No & - & 10 - 48 & 1 \\
			\hline
			$X^{33}$ & 2 & 32 & No & - & 14 - 76 & 120 \\
			\hline
			$X^{65}$ & 4 & 64 & No & - & 12 - 68 & 1 \\
			\hline
			$X^{129}$ & 2 & 32 & No & - & 14 - 81 & 660 \\
			\hline
			S\_A81  & 2 & 32 & No & - & 14 - 68 & 120 \\
			\hline 
			S\_81 & 8 & 32 & No & - & 14 - 70 & 5 \\
			\hline
			S\_82 & 8 & 64 & No & - & 11 - 48 & 2 \\
			\hline
			S\_83 & 8 & 64 & No & - & 10 - 46 & 8 \\
			\hline
			$X^3$ truncated & 4 & 32 & No & - & 14 - 70 & 80 \\
			\hline
		\end{tabular} 
	}
		\caption{Results for 8-bit S-boxes.} \label{results8} 
		
	\end{table}
	
\subsection{Over 9 bits (Table \ref{results9})}

Power maps use the field polynomial $X^9+X^4+1$.

These results are the best we obtained, but better implementations may exist. We tried to obtain one AND less, limiting the runtime to a few days, without success.

\begin{table} 
	\adjustbox{valign=t,max width=.95\columnwidth}{
	\begin{tabular}{cccc|c|c}
			\hline
			&&&&\multicolumn{1}{c}{Our implementation} & \multicolumn{1}{c}{Our timings} \\
			S-box & $\delta$ & $\mathcal{L}$ & Bij & (AND - XOR) & Config 2 \\
			&&&&& \multicolumn{1}{c}{(in \textbf{minutes})} \\
			\hline
			$\chi_9$ & 128 & 256 & Yes & 9 - 18 & $<$ 1 \\
			\hline
			$X^3$  & 2 & 32 & Yes & 19 - 94 & 5\\
			\hline
			$X^5$ & 2 & 32 & Yes & 20 - 109 & 4 \\
			\hline
			$X^9$ & 8 & 64 & Yes & 18 - 100 & 5 \\
			\hline
			$X^{65}$ & 8 & 64 & Yes & 18 - 97 & 12 \\
			\hline
			$X^3 \oplus X^5$ & 4 & 64 & No & 18 - 100 & 390 \\
			\hline
			$X^3 \oplus X^9$ & 8 & 64 & No  & 18 - 93 & 480 \\
			\hline
			$X^3 \oplus X^5 \oplus X^{17}$ & 8 & 256 & No & 18 - 96 & 4 \\
			\hline
		\end{tabular} 
	}
		\caption{Results for 9-bit S-boxes.} \label{results9}
	\end{table}

\subsection{Results obtained using the ABC tool}

We tested the ABC tool on some S-boxes. We used instructions dc2, then balance, rewrite, refactor and resub, and repeated the process 5 times to enhance the results. The corresponding results are in Table~\ref{resultsABC}. This illustrates that generic implementation tools are not fit for cryptographic purposes and justifies the need for more dedicated tools.

\begin{table} 
	\adjustbox{valign=t,max width=.95\columnwidth}{

		\begin{tabular}{c|cc|cc}
			\hline
			S-box & \multicolumn{2}{c}{Implementation (AND)} & \multicolumn{2}{c}{Timings (in sec)} \\
			& ABC & Ours & ABC & Ours \\
			\hline
			ASCON & 36 & 5 & $<$ 1 & 0.05\\
			\hline
			$X^3$ over 5 bits & 61 & 7 & $<$ 1  & $<$ 1 \\
			\hline
			$X^3$ over 6 bits & 88 & 8 & $<$ 1 & $<$ 1 \\
			\hline
			$X^3$ over 7 bits & 268 & 11 & $<$ 1 & 180\\
			\hline
			$X^3$ over 8 bits & 454 & 14 & $<$ 1 & 6900\\
			\hline
			$X^3$ over 9 bits & 676 & 19 & $<$ 1 & 300\\
			\hline
		\end{tabular} 
	}
	\caption{Results using the ABC tool.} \label{resultsABC}
\end{table}

\section{Conclusion}\label{sec:conclusions}
\noindent
We present a new tool to compute low-AND implementations of quadratic S-boxes. Our algorithm is based on precomputing 1-AND gate quadratic Boolean functions and solving a big linear system, with care about search space size and memory storage. Our tool outperforms all the existing AND-optimising tools, both in runtime and found implementations. These good runtimes are an indication of the amount of effort that would be needed to reach larger sizes. In particular, we expect that size 10 is already reachable but requires a better settup than Config. 2. For other metrics (\emph{e.g.} the ones in~\cite{DBLP:conf/fse/Stoffelen16,DBLP:journals/tcasI/FengWZPZ24}), such as Gate Equivalent Complexity or Bit-Slice Gate Complexity, our tool is suboptimal and not easily adaptable. \\
We plan to extend these results to higher-degree S-boxes, still copying what we would do by hand but with a more heuristic approach. Details of why the method presented here is not easily adaptable to higher degrees and how the problem could be tackled are given in Appendix~\ref{appendix:Limitations for higher degrees}.

\subsection*{Acknowledgements}
This work was partially supported by the French National Agency of Research under grant number ANR-22-CE39-0015.

\bibliographystyle{IEEEtran}
\bibliography{refs_MM}

\begin{thebibliography}{10}
\providecommand{\url}[1]{#1}
\csname url@samestyle\endcsname
\providecommand{\newblock}{\relax}
\providecommand{\bibinfo}[2]{#2}
\providecommand{\BIBentrySTDinterwordspacing}{\spaceskip=0pt\relax}
\providecommand{\BIBentryALTinterwordstretchfactor}{4}
\providecommand{\BIBentryALTinterwordspacing}{\spaceskip=\fontdimen2\font plus
\BIBentryALTinterwordstretchfactor\fontdimen3\font minus
  \fontdimen4\font\relax}
\providecommand{\BIBforeignlanguage}[2]{{%
\expandafter\ifx\csname l@#1\endcsname\relax
\typeout{** WARNING: IEEEtran.bst: No hyphenation pattern has been}%
\typeout{** loaded for the language `#1'. Using the pattern for}%
\typeout{** the default language instead.}%
\else
\language=\csname l@#1\endcsname
\fi
#2}}
\providecommand{\BIBdecl}{\relax}
\BIBdecl

\bibitem{DBLP:conf/crypto/IshaiSW03}
Y.~Ishai, A.~Sahai, and D.~A. Wagner, ``Private circuits: Securing hardware
  against probing attacks,'' in \emph{{CRYPTO} 2003, Santa Barbara, California,
  USA, August 17-21, 2003, Proceedings}, ser. LNCS, vol. 2729, 2003, pp.
  463--481.

\bibitem{bilginthesis}
B.~Bilgin, ``{Threshold Implementations: As Countermeasure Against Higher-Order
  Differential Power Analysis},'' Ph.D. dissertation, KU Leuven, Belgium \&
  UTwente, The Netherlands, 2015.

\bibitem{DBLP:conf/eurocrypt/BiryukovCBP03}
A.~Biryukov, C.~D. Canni{\`{e}}re, A.~Braeken, and B.~Preneel, ``A toolbox for
  cryptanalysis: Linear and affine equivalence algorithms,'' in \emph{Advances
  in Cryptology - {EUROCRYPT} 2003, International Conference on the Theory and
  Applications of Cryptographic Techniques, Warsaw, Poland, May 4-8, 2003,
  Proceedings}, ser. LNCS, E.~Biham, Ed., vol. 2656, 2003, pp. 33--50.

\bibitem{DBLP:conf/ima/DaemenR01}
J.~Daemen and V.~Rijmen, ``The wide trail design strategy,'' in
  \emph{Cryptography and Coding, 8th {IMA} International Conference,
  Cirencester, UK, December 17-19, 2001, Proceedings}, ser. LNCS, vol. 2260,
  2001, pp. 222--238.

\bibitem{DBLP:journals/tosc/BilginMDLS20}
\BIBentryALTinterwordspacing
B.~Bilgin, L.~D. Meyer, S.~Duval, I.~Levi, and F.~Standaert, ``Low {AND} depth
  and efficient inverses: a guide on s-boxes for low-latency masking,''
  \emph{{IACR} Trans. Symmetric Cryptol.}, vol. 2020, no.~1, pp. 144--184,
  2020. [Online]. Available:
  \url{https://doi.org/10.13154/tosc.v2020.i1.144-184}
\BIBentrySTDinterwordspacing

\bibitem{DBLP:conf/dagstuhl/BertoniDPA09a}
G.~Bertoni, J.~Daemen, M.~Peeters, and G.~V. Assche, ``The road from panama to
  keccak via radiogat{\'{u}}n,'' in \emph{Symmetric Cryptography, 11.01. -
  16.01.2009}, ser. Dagstuhl Seminar Proceedings, H.~Handschuh, S.~Lucks,
  B.~Preneel, and P.~Rogaway, Eds., vol. 09031, 2009.

\bibitem{DBLP:journals/joc/DobraunigEMS21}
\BIBentryALTinterwordspacing
C.~Dobraunig, M.~Eichlseder, F.~Mendel, and M.~Schl{\"{a}}ffer, ``Ascon v1.2:
  Lightweight authenticated encryption and hashing,'' \emph{J. Cryptol.},
  vol.~34, no.~3, p.~33, 2021. [Online]. Available:
  \url{https://doi.org/10.1007/s00145-021-09398-9}
\BIBentrySTDinterwordspacing

\bibitem{ullrich2011finding}
M.~Ullrich, C.~De~Canniere, S.~Indesteege, {\"O}.~K{\"u}{\c{c}}{\"u}k,
  N.~Mouha, and B.~Preneel, ``Finding optimal bitsliced implementations of
  4$\times$ 4-bit {S}-boxes,'' in \emph{SKEW 2011 Symmetric Key Encryption
  Workshop, Copenhagen, Denmark}, 2011, pp. 16--17.

\bibitem{Gla}
\BIBentryALTinterwordspacing
B.~Gladman, ``Finding efficient boolean function decompositions for the serpent
  s-boxes and their inverses,'' Accessed: 2025-06. [Online]. Available:
  \url{http://brg.a2hosted.com//oldsite/cryptography_technology/serpent/anal1.cpp}
\BIBentrySTDinterwordspacing

\bibitem{DBLP:journals/joc/BoyarMP13}
\BIBentryALTinterwordspacing
J.~Boyar, P.~Matthews, and R.~Peralta, ``Logic minimization techniques with
  applications to cryptology,'' \emph{J. Cryptol.}, vol.~26, no.~2, pp.
  280--312, 2013. [Online]. Available:
  \url{https://doi.org/10.1007/s00145-012-9124-7}
\BIBentrySTDinterwordspacing

\bibitem{DBLP:journals/tosc/JeanPST17}
\BIBentryALTinterwordspacing
J.~Jean, T.~Peyrin, S.~M. Sim, and J.~Tourteaux, ``Optimizing implementations
  of lightweight building blocks,'' \emph{{IACR} Trans. Symmetric Cryptol.},
  vol. 2017, no.~4, pp. 130--168, 2017. [Online]. Available:
  \url{https://doi.org/10.13154/tosc.v2017.i4.130-168}
\BIBentrySTDinterwordspacing

\bibitem{DBLP:journals/tosc/BaoGLS19}
\BIBentryALTinterwordspacing
Z.~Bao, J.~Guo, S.~Ling, and Y.~Sasaki, ``{PEIGEN} - a platform for evaluation,
  implementation, and generation of s-boxes,'' \emph{{IACR} Trans. Symmetric
  Cryptol.}, vol. 2019, no.~1, pp. 330--394, 2019. [Online]. Available:
  \url{https://doi.org/10.13154/tosc.v2019.i1.330-394}
\BIBentrySTDinterwordspacing

\bibitem{DBLP:conf/fse/Stoffelen16}
K.~Stoffelen, ``Optimizing {S}-box implementations for several criteria using
  {SAT} solvers,'' in \emph{Fast Software Encryption - {FSE} 2016, Bochum,
  Germany, March 20-23, 2016}, ser. LNCS, vol. 9783, 2016, pp. 140--160.

\bibitem{DBLP:journals/iacr/ZhangH23}
\BIBentryALTinterwordspacing
F.~Zhang and Z.~Huang, ``Optimizing s-box implementations using {SAT} solvers:
  Revisited,'' \emph{{IACR} Cryptol. ePrint Arch.}, p. 1721, 2023. [Online].
  Available: \url{https://eprint.iacr.org/2023/1721}
\BIBentrySTDinterwordspacing

\bibitem{DBLP:journals/tcasI/FengWZPZ24}
\BIBentryALTinterwordspacing
J.~Feng, Y.~Wei, F.~Zhang, E.~Pasalic, and Y.~Zhou, ``Novel optimized
  implementations of lightweight cryptographic s-boxes via {SAT} solvers,''
  \emph{{IEEE} Trans. Circuits Syst. {I} Regul. Pap.}, vol.~71, no.~1, pp.
  334--347, 2024. [Online]. Available:
  \url{https://doi.org/10.1109/TCSI.2023.3325559}
\BIBentrySTDinterwordspacing

\bibitem{DBLP:journals/iacr/MeyerB18}
\BIBentryALTinterwordspacing
L.~{De Meyer} and B.~Bilgin, ``Classification of balanced quadratic
  functions,'' \emph{{IACR} Cryptology ePrint Archive}, vol. 2018, p. 113,
  2018. [Online]. Available: \url{http://eprint.iacr.org/2018/113}
\BIBentrySTDinterwordspacing

\bibitem{DBLP:conf/crypto/ChariJRR99}
S.~Chari, C.~S. Jutla, J.~R. Rao, and P.~Rohatgi, ``Towards sound approaches to
  counteract power-analysis attacks,'' in \emph{{CRYPTO} '99, Santa Barbara,
  California, USA, August 15-19, 1999, Proceedings}, ser. LNCS, vol. 1666,
  1999, pp. 398--412.

\bibitem{DBLP:conf/ches/GoubinP99}
L.~Goubin and J.~Patarin, ``{DES} and differential power analysis (the
  "duplication" method),'' in \emph{Cryptographic Hardware and Embedded
  Systems, First International Workshop, CHES'99, Worcester, MA, USA, August
  12-13, 1999, Proceedings}, ser. LNCS, {\c{C}}.~K. Ko{\c{c}} and C.~Paar,
  Eds., vol. 1717, 1999, pp. 158--172.

\bibitem{DBLP:conf/tcc/MicaliR04}
S.~Micali and L.~Reyzin, ``Physically observable cryptography (extended
  abstract),'' in \emph{Theory of Cryptography, First Theory of Cryptography
  Conference, {TCC} 2004, Cambridge, MA, USA, February 19-21, 2004,
  Proceedings}, ser. LNCS, M.~Naor, Ed., vol. 2951, 2004, pp. 278--296.

\bibitem{DBLP:journals/cacm/Shamir79}
\BIBentryALTinterwordspacing
A.~Shamir, ``How to share a secret,'' \emph{Commun. {ACM}}, vol.~22, no.~11,
  pp. 612--613, 1979. [Online]. Available:
  \url{https://doi.org/10.1145/359168.359176}
\BIBentrySTDinterwordspacing

\bibitem{DBLP:conf/mark2/Blakley79}
\BIBentryALTinterwordspacing
G.~R. Blakley, ``Safeguarding cryptographic keys,'' in \emph{1979 International
  Workshop on Managing Requirements Knowledge, {MARK} 1979, New York, NY, USA,
  June 4-7, 1979}, 1979, pp. 313--318. [Online]. Available:
  \url{https://doi.org/10.1109/MARK.1979.8817296}
\BIBentrySTDinterwordspacing

\bibitem{DBLP:journals/jce/BozilovKN22}
\BIBentryALTinterwordspacing
D.~Bozilov, M.~Knezevic, and V.~Nikov, ``Optimized threshold implementations:
  securing cryptographic accelerators for low-energy and low-latency
  applications,'' \emph{J. Cryptogr. Eng.}, vol.~12, no.~1, pp. 15--51, 2022.
  [Online]. Available: \url{https://doi.org/10.1007/s13389-021-00276-5}
\BIBentrySTDinterwordspacing

\bibitem{DBLP:conf/ccs/BartheBDFGSZ16}
\BIBentryALTinterwordspacing
G.~Barthe, S.~Bela{\"{\i}}d, F.~Dupressoir, P.~Fouque, B.~Gr{\'{e}}goire,
  P.~Strub, and R.~Zucchini, ``Strong non-interference and type-directed
  higher-order masking,'' in \emph{Proceedings of the 2016 {ACM} {SIGSAC}
  Conference on Computer and Communications Security, Vienna, Austria, October
  24-28, 2016}, 2016, pp. 116--129. [Online]. Available:
  \url{https://doi.org/10.1145/2976749.2978427}
\BIBentrySTDinterwordspacing

\bibitem{DBLP:journals/tches/GoudarziPRV21}
\BIBentryALTinterwordspacing
D.~Goudarzi, T.~Prest, M.~Rivain, and D.~Vergnaud, ``Probing security through
  input-output separation and revisited quasilinear masking,'' \emph{{IACR}
  Trans. Cryptogr. Hardw. Embed. Syst.}, vol. 2021, no.~3, pp. 599--640, 2021.
  [Online]. Available: \url{https://doi.org/10.46586/tches.v2021.i3.599-640}
\BIBentrySTDinterwordspacing

\bibitem{DBLP:journals/iacr/BronchainCS21}
\BIBentryALTinterwordspacing
O.~Bronchain, G.~Cassiers, and F.~Standaert, ``Give me 5 minutes: Attacking
  {ASCAD} with a single side-channel trace,'' \emph{{IACR} Cryptol. ePrint
  Arch.}, p. 817, 2021. [Online]. Available:
  \url{https://eprint.iacr.org/2021/817}
\BIBentrySTDinterwordspacing

\bibitem{DBLP:journals/tosc/BelliziaBBCDGLL20}
\BIBentryALTinterwordspacing
D.~Bellizia, F.~Berti, O.~Bronchain, G.~Cassiers, S.~Duval, C.~Guo, G.~Leander,
  G.~Leurent, I.~Levi, C.~Momin, O.~Pereira, T.~Peters, F.~Standaert,
  B.~Udvarhelyi, and F.~Wiemer, ``Spook: Sponge-based leakage-resistant
  authenticated encryption with a masked tweakable block cipher,'' \emph{{IACR}
  Trans. Symmetric Cryptol.}, vol. 2020, no.~{S1}, pp. 295--349, 2020.
  [Online]. Available: \url{https://doi.org/10.13154/tosc.v2020.iS1.295-349}
\BIBentrySTDinterwordspacing

\bibitem{DBLP:conf/ches/JournaultS17}
\BIBentryALTinterwordspacing
A.~Journault and F.~Standaert, ``Very high order masking: Efficient
  implementation and security evaluation,'' in \emph{Cryptographic Hardware and
  Embedded Systems - {CHES} 2017 - 19th International Conference, Taipei,
  Taiwan, September 25-28, 2017, Proceedings}, ser. LNCS, W.~Fischer and
  N.~Homma, Eds., vol. 10529, 2017, pp. 623--643. [Online]. Available:
  \url{https://doi.org/10.1007/978-3-319-66787-4\_30}
\BIBentrySTDinterwordspacing

\bibitem{DBLP:conf/eurocrypt/GoudarziR17}
\BIBentryALTinterwordspacing
D.~Goudarzi and M.~Rivain, ``How fast can higher-order masking be in
  software?'' in \emph{Advances in Cryptology - {EUROCRYPT} 2017 - 36th Annual
  International Conference on the Theory and Applications of Cryptographic
  Techniques, Paris, France, April 30 - May 4, 2017, Proceedings, Part {I}},
  ser. LNCS, J.~Coron and J.~B. Nielsen, Eds., vol. 10210, 2017, pp. 567--597.
  [Online]. Available: \url{https://doi.org/10.1007/978-3-319-56620-7\_20}
\BIBentrySTDinterwordspacing

\bibitem{DBLP:books/sp/DaemenR02}
\BIBentryALTinterwordspacing
J.~Daemen and V.~Rijmen, \emph{The Design of Rijndael: {AES} - The Advanced
  Encryption Standard}, ser. Information Security and Cryptography, 2002.
  [Online]. Available: \url{https://doi.org/10.1007/978-3-662-04722-4}
\BIBentrySTDinterwordspacing

\bibitem{DBLP:conf/ches/BogdanovKLPPRSV07}
A.~Bogdanov, L.~R. Knudsen, G.~Leander, C.~Paar, A.~Poschmann, M.~J.~B.
  Robshaw, Y.~Seurin, and C.~Vikkelsoe, ``{PRESENT:} an ultra-lightweight block
  cipher,'' in \emph{Cryptographic Hardware and Embedded Systems - {CHES} 2007,
  Vienna, Austria, September 10-13, 2007, Proceedings}, ser. LNCS, vol. 4727,
  2007, pp. 450--466.

\bibitem{DBLP:journals/tssc/HartNR68}
\BIBentryALTinterwordspacing
P.~E. Hart, N.~J. Nilsson, and B.~Raphael, ``A formal basis for the heuristic
  determination of minimum cost paths,'' \emph{{IEEE} Trans. Syst. Sci.
  Cybern.}, vol.~4, no.~2, pp. 100--107, 1968. [Online]. Available:
  \url{https://doi.org/10.1109/TSSC.1968.300136}
\BIBentrySTDinterwordspacing

\bibitem{moustakas1990heuristic}
C.~Moustakas, \emph{Heuristic research: Design, methodology, and applications},
  1990.

\bibitem{DBLP:conf/space/DaemenMM21}
\BIBentryALTinterwordspacing
J.~Daemen, A.~Mehrdad, and S.~Mella, ``Computing the distribution of
  differentials over the non-linear mapping {\(\chi\)},'' in \emph{Security,
  Privacy, and Applied Cryptography Engineering - 11th International
  Conference, {SPACE} 2021, Kolkata, India, December 10-13, 2021, Proceedings},
  ser. LNCS, L.~Batina, S.~Picek, and M.~Mondal, Eds., vol. 13162, 2021, pp.
  3--21. [Online]. Available:
  \url{https://doi.org/10.1007/978-3-030-95085-9\_1}
\BIBentrySTDinterwordspacing

\bibitem{DBLP:journals/dcc/BrinkmannL08}
\BIBentryALTinterwordspacing
M.~Brinkmann and G.~Leander, ``On the classification of {APN} functions up to
  dimension five,'' \emph{Des. Codes Cryptography}, vol.~49, no. 1-3, pp.
  273--288, 2008. [Online]. Available:
  \url{https://doi.org/10.1007/s10623-008-9194-6}
\BIBentrySTDinterwordspacing

\bibitem{DBLP:journals/tit/BeierleL22}
\BIBentryALTinterwordspacing
C.~Beierle and G.~Leander, ``New instances of quadratic {APN} functions,''
  \emph{{IEEE} Trans. Inf. Theory}, vol.~68, no.~1, pp. 670--678, 2022.
  [Online]. Available: \url{https://doi.org/10.1109/TIT.2021.3120698}
\BIBentrySTDinterwordspacing

\end{thebibliography}

\begin{IEEEbiography} [{\includegraphics[width=1.7cm]{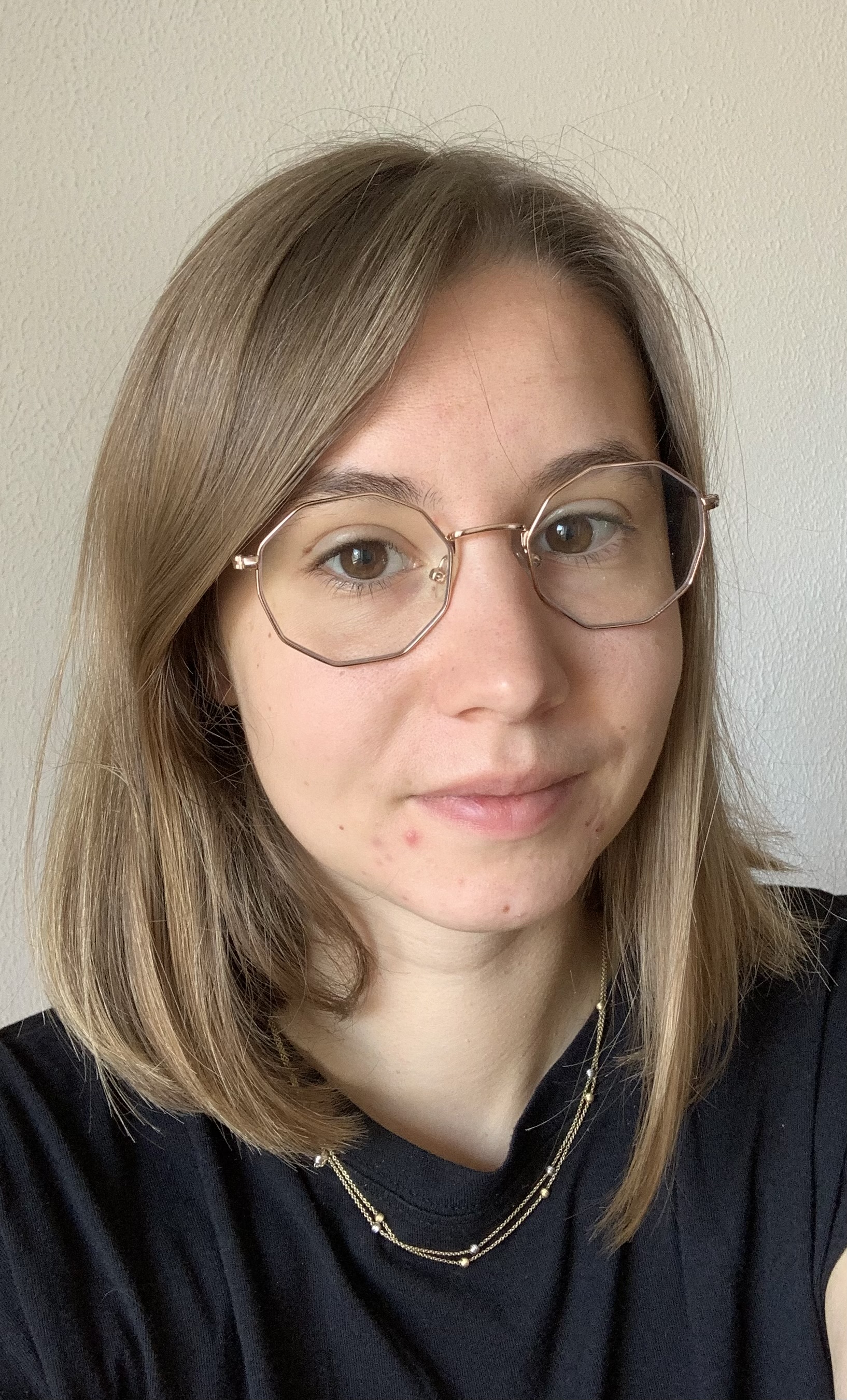}}]
	{Marie Bolzer} is currently pursuing the Ph.D. degree at Universit\'e de Lorraine and at the LORIA Lab. Her research interests include algorithmics and the development of automated tools for constructing and analysing symmetric-key cryptography components.
\end{IEEEbiography}

\begin{IEEEbiography} [{\includegraphics[width=1.7cm]{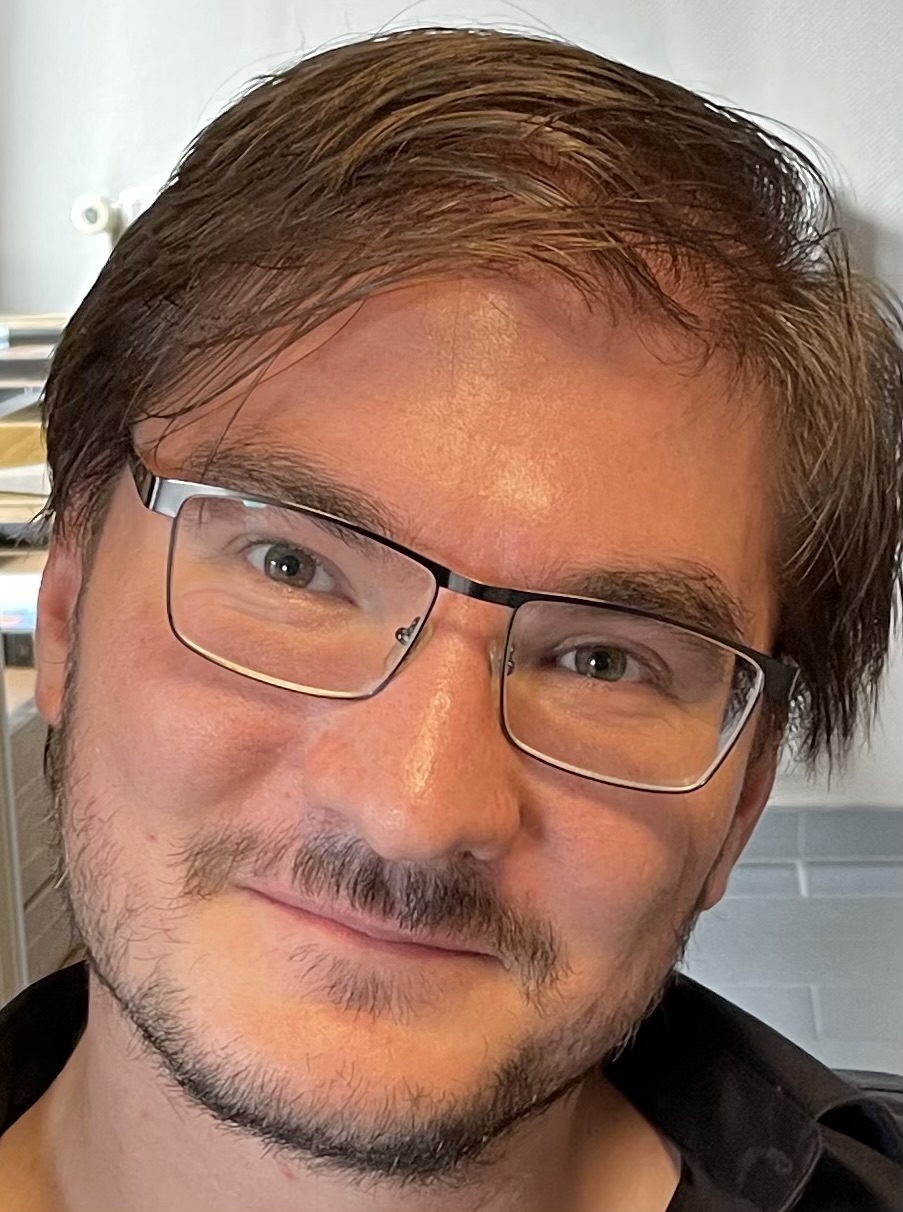}}]
	{Sébastien Duval} received the Ph.D. degree in computer science from Sorbonne Universit\'e in 2018. Since 2021, he is an associate professor at Universit\'e de Lorraine and at the LORIA lab. His research interests include design of lightweight symmetric-key cryptography and real-world security of cryptography.
\end{IEEEbiography}

\begin{IEEEbiography}[{\includegraphics[width=1.7cm]{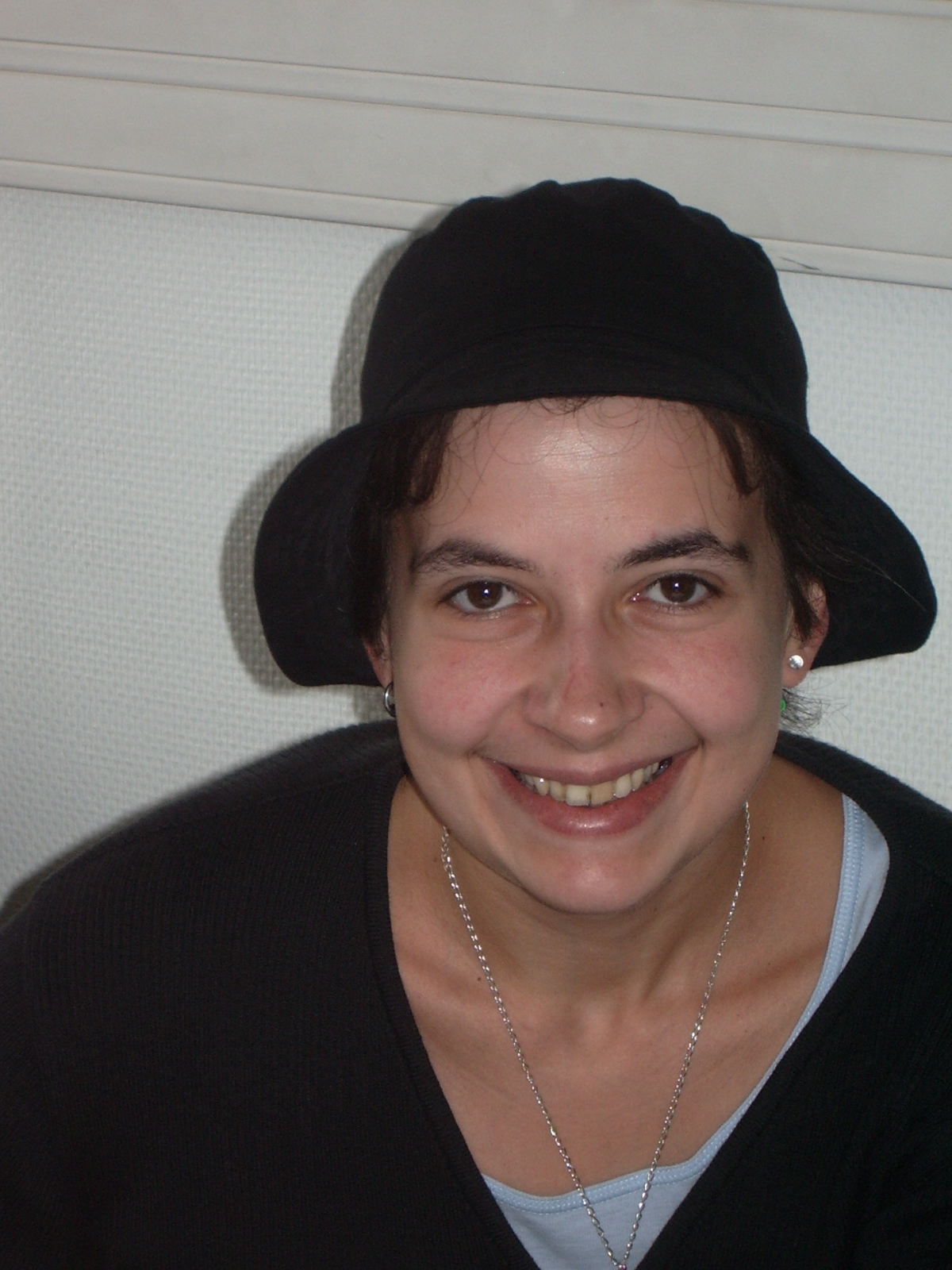}}]
	{Marine Minier} received the Ph.D. degree in 2002, from the Universit\'e de Limoges and the French Habilitation from the Universit\'e de Lyon in 2012. In 2005, she joined the INSA de Lyon and the CITI Laboratory, as an Assistant Professor. Since 2016, she is professor at Universit\'e de Lorraine and at the LORIA Lab. Her research interests include Symmetric Key Cryptography and Security in WSNs.
\end{IEEEbiography}

\appendices

\section{Limitations for higher degrees}\label{appendix:Limitations for higher degrees}
\noindent

To use a method similar to the one presented here for a degree $d>2$, we would like to reduce the problem to a linear system as in the quadratic case. We would thus like to obtain any function of degree $d$ as a XOR-sum of elements of $\textsf{Set\_op}$. \\
Let us start with $d=3$ and denote $L_n$ the set of linear functions on $n$ bits. To obtain any function of degree 3 on $n$ bits, we can start by considering the XOR-sum of elements of the form $l_1 \times l_2 \times l_3$, with $l_i \in L_n$.

\begin{remark}
	Each of the elements obtained in this way costs 2 AND unless one of the $l_i$ is equal to 1, in which case these elements correspond to those present in $\textsf{Set\_op}$ and cost 1 AND. Note that, unlike the quadratic case where all operations require 1 AND, the elements no longer have the same cost. As a result, the function \textsf{test\_xor} must be adapted to take this cost into account when adding operations to \textsf{Op\_selec}.
\end{remark}

We denote this new set $\textsf{Set\_op}_3$. The number of elements in this new set is much larger but could remain reasonable for small sizes. (column 3 of Table \ref{Cardinals}).
However, to optimise the multiplicative complexity, we should consider other types of decomposition, as shown in the example below.

\begin{example}
	Consider $y = x_0x_1x_3 \oplus x_2x_3$. \\
	We want to express $y$ as a XOR-sum of elements of $\textsf{Set\_op}_3$. $y$ is not in $\textsf{Set\_op}_3$, so we want to know if a XOR-sum of 2 elements of $\textsf{Set\_op}_3$ is equal to $y$, the implementation obtained will then cost strictly more than 2 AND.
	We thus obtain $y = c_1 \oplus c_2$, with $c_1 = x_0x_1x_3 \in \textsf{Set\_op}_3$ and $c_2 = x_2x_3 \in \textsf{Set\_op}_3$. The resulting implementation requires $2 + 1 = 3$ AND. \\
	However, we can see that $y = ((x_0 \times x_1) \oplus x_2) \times x_3$ and that this decomposition only costs 2 AND. Decompositions costing only 2 AND are therefore not all achievable with only elements of the form $l_1 \times l_2 \times l_3$, with $l_i \in L_n$. \\
\end{example}

Let us therefore also consider elements of the form $(l_1 \times l_2 + l_3)\times l_4$ , with $l_i \in L_n$. By adding these decompositions to $\textsf{Set\_op}_3$ we obtain the cardinals of the 4th column (v2) of Table \ref{Cardinals}. \\
In the same way, elements of the form $((l_1 \times l_2)\oplus (l_3 \times l_4) \oplus l_5) \times l_6$, with $l_i \in L_n$ could also be added as they cost 3 AND and would only be obtained by decompositions of the form $(l_1 \times l_2 \times l_6)\oplus ((l_3 \times l_4) \oplus l_5) \times l_6$ which cost 4 AND. We would then obtain the cardinals of the 5th column (v3) of Table \ref{Cardinals}.

\begin{table}[h]
	\centering
	\begin{tabular}{c|c|c|c|c}
		Nb bits & \#\textsf{Set\_op} & \#$\textsf{Set\_op}_3$ v1 & \#$\textsf{Set\_op}_3$ v2 & \#$\textsf{Set\_op}_3$ v3 \\
		\hline
		4 & 35 & 141 & 561 & 561 \\
		5 & 155 & 1241 & 14 261 & 28 149 \\
		6 & 651 &  10 417 & 283 387 & 2 033 725 \\
	\end{tabular} 
	\caption{Cardinals of operation sets for degree 3}
	\label{Cardinals}
\end{table}

In summary, we cannot keep this efficient approach for degree $d>2$, as it requires storing too many elements. A sensible approach might be to use the formulas given above to decompose a given function, as these formulas are possible factorisations, with their associate cost in terms of AND gates.

\end{document}